\documentclass[12pt]{amsart}
\usepackage{amsmath}
\usepackage{amsbsy}
\usepackage{graphicx,amsmath,amssymb,fullpage,amsfonts,bbm,bbold}
\usepackage{bbm} 
\usepackage{setspace}
\usepackage{color} 
\usepackage{enumitem}
 \usepackage{dsfont}

\usepackage{tikz}
\usetikzlibrary{arrows}
\usetikzlibrary{positioning}
\newdimen\nodeDist
\newcommand{\X}{\mbox{x}}

\newcommand{\Ts}{\mathcal{T}}

\newcommand{\N}{\mbox{{\small\textsc{N}}}}

\newcommand{\link}{F^{-1}}

\usepackage{bm}
\usepackage{natbib}
\title{A {B}ayesian partial identification approach to inferring the prevalence of accounting misconduct} 
\author{P. Richard Hahn$^1$}
\thanks{1. Booth School of Business, University of Chicago}
\author{Jared S. Murray$^2$}
\thanks{2. Department of Statistics, Carnegie Mellon University}
\author{Ioanna Manolopoulou$^3$}
\thanks{3.  Department of Statistical Science, University College of London}
\thanks{The first author thanks the Booth School of Business for supporting this research.}
\thanks{The second author was supported in part by the National Science Foundation under grant numbers SES-11-31897 and SES-1130706. Any opinions, findings, and conclusions or recommendations 
expressed in this material are those of the author(s) and do not necessarily
 reflect the views of the National Science Foundation.}
\date{}

\begin{document}
\maketitle
\singlespace 
\begin{abstract}
This paper describes the use of flexible Bayesian regression models for estimating a partially identified probability function. Our approach permits efficient sensitivity analysis concerning the posterior impact of priors on the partially identified component of the regression model.  The new methodology is illustrated on an important problem where only partially observed data is available-- inferring the prevalence of accounting misconduct among publicly traded U.S. businesses.
\end{abstract}

\doublespace
\section{Introduction}
This paper develops an approach for estimating partially identified parameters in nonlinear regression settings. Our approach is based on a  decomposition of the probability function into an identified  and a partially identified component \citep{Kadane1975}.  This representation permits us to employ flexible (nonlinear) models when inferring the identified component; in our applications we utilize Bayesian tree-based priors for the regression functions \citep{chipman2010bart, Hill2012}.  For the partially identified portion of the model, informative priors are crucial, so checking the sensitivity of posterior inferences to model specification is vital. In our proposed framework, this sensitivity analysis is straightforward, and may be conducted under many different models for the partially identified parameters using only one set of samples from the marginal posterior of the identified parameters. 

Our motivating application comes from the corporate accounting literature, where there is substantial interest in determining what fraction of U.S. firms engage in financial misconduct (such as misstated earnings);  e.g. \cite{Dyck2013}.  Inferring the prevalence of misconduct is complicated by an inherent partial observability---not all cases of misconduct are discovered. Any treatment of this problem will therefore need to analyze how company attributes impact the probability of misconduct being discovered in addition to the probability of the misconduct itself taking place.  

As further evidence of the generality of our approach, we also include a reanalysis of a published dataset (from a broken randomized encouragement study of flu vaccine)  in the supplementary material.

The remainder of this section collects necessary background material, providing an overview of the concept of partial identification (specifically its treatment from a Bayesian perspective) and describing the empirical data we will analyze.  Section \ref{Kadane} lays out our framework more generally and fixes notation.  Section \ref{empirical} describes the results of our data analysis.
 
\subsection{Partial identification}
A statistical model $p(y\mid\tau)$ indexed by a parameter $\tau\in\Ts$ is said to be {\em identifiable} or {\em identified} if parameter values correspond uniquely to distinct probability distributions over observables. That is, $p(y\mid\tau)=p(y\mid\tau')$ for all $y$ if and only if $\tau=\tau'$. A model that is not identified is simply referred to as unidentified. 
The importance of identifiability as a modeling concern has its earliest roots in econometrics, the term first being introduced in \cite{Koopmans1949}.  Other seminal references include \cite{haavelmo1943, haavelmo1944} and \cite{KoopmansReiersol1950}.  Unidentifiability arises naturally in econometric analysis of observational data as a byproduct of imperfect measurement and/or various data censoring mechanisms. 
The Bayesian perspective on identifiability has been comprehensively reviewed in \cite{Aldrich2002} and more recently in \cite{MartinGonzalez2010}. 

The notion of {\em partial identifiability} or {\em partial identification} of parameters expands the concept of identification to consider cases of partial learning. A more general definition of identifiability is $p(y\mid\tau)=p(y\mid\tau')$ if and only if $t(\tau)=t(\tau')$ for some non-constant function $t$; here $t$ is an ``identifying function" in the terminology of \cite{Kadane1975}. When $t$ is one-to-one, we recover the traditional definition of identifiability, or {\em point identification}. When the $t$ with the finest preimage satisfying this condition is many-to-one, the model is partially identified --- the intuition being that asymptotically we can only isolate the value of $t(\tau)$ consistent with the data, which will correspond to a proper subset of $\Ts$ with more than one element. For this reason, and in contrast to point identification, it is common to talk of {\em set identification}. For a more rigorous exposition of the theory of functional identification, refer to \cite{Kadane1975}.
 
Early examples of the partial identification concept include \cite{frisch1934statistical}, \cite{frechet1951tableaux} and \cite{duncan1953alternative}.  In recent years, interest in partial identification has accelerated; an excellent recent review article is \cite{Tamer2010} which includes comprehensive citations. See also the book-length treatments by Manski  \citep{manski1995identification,manski2003partial,manski2007identification}.
%
Recent contributions from a Bayesian perspective have focused primarily on asymptotic properties of the posterior distribution over partially identified parameters, notably \cite{gustafson2005model} and \cite{gustafson2010bayesian}. \cite{moon2012bayesian} examine asymptotic discrepancies between Bayesian credible regions and frequentist confidence sets for set-identified parameters.  \cite{florens2011bayesian} consider a theoretical framework for studying posteriors of partially identified parameters in nonparametric models.
\cite{kline2013default} develop large sample approximations of posterior probabilities that particular parameter values lie in the identified set without reference to a prior on the partially identified parameter.

Our approach differs from these recent contributions in three ways.  One, it is tailored to a nonlinear regression setting with possibly many predictors and complicated inter-relationships; most of the recent literature considers much simpler examples, often without any covariates.  Two, our focus is on practical methods for making inferences on parameters of interest with finite samples; most of the recent literature has focused on theoretical and specifically large-sample issues.  Three, we introduce an efficient computational scheme for sensitivity analysis, an issue which has received relatively little attention in 
the literature.  Most previous work focuses on wholly unidentified parameters and typically requires multiple iterations of model fitting; see e.g. \cite{McCandless2007, Molitor2009} and \cite{McCandless2012} in the context of causal inference/observational data analysis, and \cite{Daniels2008} and chapter 15 of \cite{little1987statistical} for extensive reviews in missing data problems.

\subsection{Application: inferring the prevalence of accounting misconduct}
Since 1982, the United States Securities and Exchange Commission (SEC) has released public notices called Accounting and Auditing Enforcement Releases, or AAERs.  AAERs are financial reports ``related to enforcement actions concerning civil lawsuits brought by the Commission in federal court and notices and orders concerning the institution and/or settlement of administrative proceedings"  \citep{SECwebsite}. Informally, AAERs comprise a list of publicly traded firms that the SEC has cited for misconduct in one form or another.  

For brevity, we adopt the nomenclature ``cheating'' and ``caught'', with the understanding that ``cheating" is operationally defined as any accounting anomaly that would lead to an AAER being issued, were it explicitly brought to the SEC's attention.  This interpretation entails that no ``caught" firms did not ``cheat", by definition.

Our goal is to provide an estimate of the prevalence of accounting misconduct in the U.S. economy, defined as all actual (caught) and potential (uncaught) AAERs.  Predicting which companies are likely to cheat, on the basis of observable firm characteristics, is complicated by the fact that there are potentially many instances of misconduct of which the SEC is unaware.  Thus, we do not directly observe which firms cheat, but merely the subset of cheating firms that were {\it caught} doing so.  A naive regression analysis would therefore only speak to the question of which attributes are predictive of getting caught cheating.  To complete the analysis, one must incorporate knowledge or conjectures concerning the impact firm attributes have on the likelihood of misconduct being discovered.

\section{Partially observable binary response vectors}\label{Kadane}
Problems with a similar structure to our accounting application appear in the literature under the heading of ``partially observed binary data".  Regression models for such data have been studied in many different fields, going by various names.  For example, \cite{Lancaster1996} considers the case where the observation model is covariate independent under the name ``contaminated case-control", building on \cite{prentice1979logistic}.  \cite{poirier1980partial} studies such data under the rubric of ``partially observed bivariate probit models", building on the work of \cite{Heckman1976,Heckman1978,Heckman1979}.  Our analysis is similar to the approach taken in \cite{Wang2013}, which adapts the bivariate probit model of \cite{poirier1980partial} for the securities fraud problem.
 
 
 Whereas these earlier references considered particular parametric models, such as the probit model, and studied identification conditions in that setting, we proceed in the more generic setting of nonlinear regression models, which leads to partially identified parameters.  Our approach will be to confront this partial identification with informative priors.

\subsection{Reparametrization in partially identified models}
As in \cite{Dawid1979,GelfandSahu1999} and \cite{gustafson2005model}, we will work with a reparameterization of $\tau$ into an identified component $\phi$ and an unidentified component $(\theta, \eta)$. We write the unidentified component as two parameters:  $\theta$, which appears in our estimand of interest, and $\eta$, which is a nuisance parameter. We will refer to $\phi$, $\phi(\X)$ or $\phi_{\X}$ (respectively, $\theta$, $\theta(\X)$ or $\theta_{\X}$ and $\eta$, $\eta(\X)$ or $\eta_{\X}$) depending on context. We will use $\phi$ when the dependence on $\X$ is inessential, we will use $\phi(\X)$ to emphasize that $\phi$ is a function of $\X$, and we will use $\phi_{\X}$ to refer to point-wise evaluations of $\phi(\X)$.

The joint distribution over data and parameters in a partially identified model can be written as 
 \begin{equation}
\begin{split}
\pi(\eta, \theta, \phi, y) & = f(y \mid \eta, \theta, \phi)\pi(\eta, \theta, \phi),\\
& = f(y \mid \phi)\pi(\eta, \theta, \phi),\\
& = f(y \mid \phi)\pi(\phi)\pi(\eta, \theta \mid \phi),\\
\end{split}
\end{equation}
where the conditional independence implied in moving from the first line to the second line constitutes a definition of partial identification. It follows that the joint posterior distribution of the identified component $\phi$ and the unidentified component $(\theta, \eta)$ can be written as
 \begin{equation}\label{composition}
\begin{split}
\pi(\eta, \theta, \phi \mid y) &= \frac{f(y \mid \phi)\pi(\phi)\pi(\eta, \theta \mid \phi)}{f(y)}\\
& = \frac{f(y \mid \phi)\pi(\phi)}{f(y)}\pi(\eta, \theta \mid \phi)\\
& = \pi(\phi \mid y)\pi(\eta, \theta \mid \phi).
\end{split}
\end{equation}
Theorem 5 of \cite{Kadane1975} shows rigorously that the parameter space of any model can be decomposed in this way.  Essentially, there are three cases to consider.  If the model is fully identified, then $(\eta, \theta)$ is empty, meaning it is a degenerate/constant random variable. When the model has fully unidentified elements, the support of $\pi(\eta, \theta \mid \phi)$ does not depend on $\phi$; the data inform about $(\eta, \theta)$ only via the presumed prior dependence represented in the choice of $\pi(\eta, \theta \mid \phi)$. In the partially identified case, which we focus on here, $\pi(\eta, \theta \mid \phi)$ has support restrictions that do depend on $\phi$;  we will denote this $\phi$-dependent support by $\Omega(\phi)$.

Our approach will be to directly specify $\pi(\eta, \theta \mid \phi)$ with support $\Omega(\phi)$.  The tractability of our approach depends on being able to construct $\pi(\eta, \theta \mid \phi)$ in a way that permits convenient sampling. Our empirical analysis (Section \ref{cparm}) provides a specific example where this approach is successful (with additional computational details in the  appendix).

Although this type of reparameterization is not novel, it has been employed in the literature mainly as a way to study large sample behavior of posterior distributions in partially identified models; see, for example, \cite{gustafson2005model,gustafson2010bayesian} and \cite{florens2011bayesian}. Most authors suggest using priors on $\tau$ to induce the priors on $(\phi, \theta)$, or simply elect to do indirect inference for $\theta$ using only $\phi$ by computing posteriors over $\Omega(\phi)$.  We consider a different approach, working directly in the $(\phi, \theta)$ parameterization and specifying priors as $\pi(\theta\mid\phi)\pi(\phi)$. This allows us to be fully nonparametric where the data are informative (i.e., on $\phi$), and exploit the conditional independence in \eqref{composition} to conduct computationally efficient sensitivity analysis. 

\subsection{Partially observed multivariate binary regression models}\label{sec:binreg}
The general structure of the problem is as follows: The complete data consist of binary vectors $U=(U_1,\dots U_k)$, of which only certain subsets are simultaneously observable. Interest is in some functional $p(\X)$ of the entire joint distribution $p(U_1,\dots U_k\mid \X)$. Due to the partial observability, $p(\X)$ must be reconstructed from an identified function $\phi(\X)$ and an unidentified function $\theta(\X)$.

Specializing to our corporate accounting analysis, let $Z_i$ indicate ``cheating" in firm-year $i$ and let $W_i$ indicate ``getting caught" in firm-year $i$, ($U_1 = W$ and $U_2 = Z$). We assume that with some probability, cheaters get caught, but that there are no firms who get caught when they are not cheating (this is consistent with our operational definition of ``cheating"). Additonally, the data are ``presence-only'' in that we have no confirmation that any given firm is certainly non-cheating.  

The parameter of interest is the marginal firm-year probability of cheating, 
\begin{equation}
p(\X)=\Pr(Z =1\mid \X) = \frac{\Pr(W=1, Z=1\mid \X)}{\Pr(W=1\mid Z=1, \X)}
\end{equation}
from which we may determine the overall prevalence of cheating across all firms as
\begin{equation}\label{alpha}
\alpha \equiv \displaystyle \sum_i \mbox{Pr}(Z_i=1 \mid \X_i) = \sum_i \frac{\mbox{Pr}(Z_i=1, W_i = 1 \mid \X_i)}{\mbox{Pr}(W_i = 1 \mid Z_i = 1, \X_i)}.
\end{equation}

Equivalently, for each firm-year we observe $Y_i \equiv Z_iW_i$ instead of $(Z_i, W_i)$, where $Y_i$ indicating whether a firm received an AAER (cheated and got caught), giving
\begin{equation}
\phi(\X) = \mbox{Pr}(Z=1, W = 1 \mid \X) = \mbox{Pr}(Y  = 1 \mid \X); \;\;\; \theta(\X)  = \mbox{Pr}(W = 1 \mid Z = 1, \X).
\end{equation} 

As $\phi(\X)$ is simply the (conditional) probability of the observed binary data $Y$, it is point identified.  In our application, we estimate $\phi(\X)$ using the BART model described in the Appendix. BART has been shown empirically to be an excellent default nonlinear regression method, with a demonstrated ability to handle many noise variables and strong nonlinearities  \citep{chipman2010bart, Hill2012}. 

The partial identification of $p(\X)$ arises simply because $0 \leq p(\X) \leq 1$.  Given $\phi(\X)$, the posterior on $p(\X)$ is defined by the prior over $\theta(\X)$, truncated to regions satisfying 
\begin{equation}\label{partial_ineq}
0 \leq \frac{\phi(\X)}{\theta(\X)} \leq 1,
\end{equation}
for all $\X \in \mathcal{X}$. In other words,  in our applied setting, $\Omega(\phi(\X)) = \{\theta(\X) \mid \phi(\X) \leq \theta(\X) \}$.



\subsection{A Gaussian process model for the partially identified regression}

Furnishing prior information regarding $\theta(\X)$ in a predictor-dependent manner strongly motivates the use of simple parametric models.  For starters, consider the construction
\begin{equation}
\link\{\theta(\X)\} = h(\X)\beta,\label{eq:theta-1} 
\end{equation}
where $F^{-1}$ is a link function and $h$ denotes some transformation or subset of the covariate vector.  A prior over $\theta(\X)$ is induced by a prior over $\beta$.  A chief difficulty with this type of specification is that nonlinear (possibly discontinuous) regression models for $\phi(\X)$ impose complex support  restrictions on $\beta$ --- indeed, some samples from the posterior for $\phi(\X)$ may contradict the model for $\theta(\X)$ entirely, meaning that they imply a set of bounds for $\theta(\X)$ such that no feasible $\beta$ exists. 

To address this problem, we expand the prior over $\theta(\X)$ to acknowledge that \eqref{eq:theta-1} is only a guess as to the form of the regression function. Specifically, we \emph{center} our model for $\theta(\X)$ at \eqref{eq:theta-1} by assuming that
\begin{equation}
\link\{\theta(\X)\} \mid \phi(\X), \eta(\X) \sim \mathcal{GP}\left(h(\X)\beta,  \Sigma_{\mathcal{X}}\right)\mathbb{1}((\theta(\X), \eta(\X)) \in\Omega\{\phi(\X)\})\label{eq:theta-2}
\end{equation}
for all $\X\in \mathcal{X}$, where $\mathcal{GP}(m, \Sigma_{\mathcal{X}})$ is a Gaussian process with mean $m$ and covariance $\Sigma_{\mathcal{X}}$. In this set-up, $\beta$ may be fixed or given a prior distribution, in which case we equate $\eta := \beta$. 

For the purpose of computing $\alpha$ as in (\ref{alpha}), the prior \eqref{eq:theta-2} applied to the observed data points reduces to a multivariate normal prior (conditional on $\beta$) truncated according to (\ref{partial_ineq}).  In our empirical analysis, we choose $\Sigma_{\mathcal{X}} = \sigma^2 \mbox{I}$.  Note that this independence is in terms of design points, not in terms of observations: $\theta_{\X_i} = \theta_{\X_j}$ if $\X_i = \X_j$, even if $i \neq j$. That is, draws from the prior \eqref{eq:theta-2} are non-smooth functions, as distinct from observation-level errors. Under this simplification, sampling from $\left(\theta_{\X}\mid \beta, \phi_{\X}\right)$ reduces to drawing samples from independent truncated univariate normal distributions.  The more complicated case of a smooth Gaussian process prior (e.g., one with a squared exponential covariance function) would require draws from a multivariate truncated normal distribution; reasonably efficient Hamiltonian Monte Carlo algorithms have been developed for this task \citep{HMC} but we do not pursue this direction here.

\subsection{Simulated example}\label{sec:ex}
Before turning to the subject-specific prior we developed for our empirical application, it is instructive to consider how our approach performs in a situation where we know what the right answers ought to be. To this end, we consider a simulated data set based on the competing approach of \cite{Wang2013}, which also analyzes the SEC data.

\cite{Wang2013} builds off \cite{poirier1980partial}, which considers a latent Gaussian utility formulation of the bivariate probit model:
\begin{equation}\label{biprobit}
\begin{pmatrix} Z^*\\ W^* \end{pmatrix} \sim \N(\mu(\X), \mbox{C}_{\rho})
\end{equation}
where $\mu(\X) = (\gamma_0 + \X^t\gamma, \beta_0 + \X^t\beta)$ and $\mbox{C}_{\rho}$ is a 2-by-2 correlation matrix with correlation parameter $\rho$.  The observed bivariate binary data is then $Z := \mathbb{1}(Z^* > 0)$ and $W := \mathbb{1}(W^* > 0)$. Specifically, \cite{poirier1980partial} establishes that (subject to certain exclusion restrictions) the parameters of the model  ($\gamma$, $\beta$ and the correlation parameter $\rho$) are still identified even when only $Y := ZW$ is observed. \cite{Wang2013} proposes to leverage this result, while deviating from the latent utility formulation.  In particular, despite making a  ``no false positives" assumption (as we do here), \cite{Wang2013} continues to equate $\Pr(Z^* > 0 \mid \X)$ with the probability of cheating, which corresponds to the somewhat arbitrary model:
\begin{equation}\label{wang}
\begin{split}
\Pr(\textrm{cheat, caught})&= \Phi(\gamma_0  + \X^t\gamma, \beta_0  + \X^t \beta, \rho)\\
\Pr(\textrm{cheat, not caught})&=  \Phi(\gamma_0  + \X^t\gamma) - \Phi(\gamma_0  + \X^t\gamma, \beta_0 + \X^t \beta, \rho)\\
\Pr(\textrm{not cheat, caught})&=0\\
\Pr(\textrm{not cheat, not caught})&= 1 - \Phi(\gamma_0 + \X^t\gamma).
\end{split}
\end{equation}
In other words, \cite{Wang2013} identifies $\gamma$ from the first equation above, invoking the result of \cite{poirier1980partial}, and then proceeds to interpret $\gamma$ as the parameter from a bivariate probit model {\em without} the no false positives assumption. While there is nothing formally wrong with this model, it would seem to defy justification.

All the same, if (\ref{wang}) is in fact the correct model, it is instructive to observe what our approach gives up to it.  Conversely, if (\ref{wang}) is used in a misspecified setting, how do its results compare to ours?  To investigate, we simulated $n = 2000$ observations from the following two models.  First, we generated data according to (\ref{wang}) by drawing $Y := WZ$ with $(W, Z)$ coming from a bivariate probit model with $\gamma_0 = \beta_0 = -1/2$, $\gamma = (-1, 3/4, 0)$, $\beta = (-3/4,0,-1/2)$, and $\rho = 1/2$, with $\X_1$ drawn from a $\mbox{Uniform}(-\pi/2, \pi/2)$ distribution, and  $\X_2$ and $\X_3$ drawn from a $\mbox{Uniform}(-3\pi/2, 3\pi/2)$ distribution (independently).  This specification of $\gamma$ and $\beta$ satisfies the exclusion restriction of \cite{poirier1980partial}, in that distinct predictor variables are omitted from each linear equation in the probit mean function.  

\begin{figure} 
\begin{center}
\includegraphics[width=3.25in]{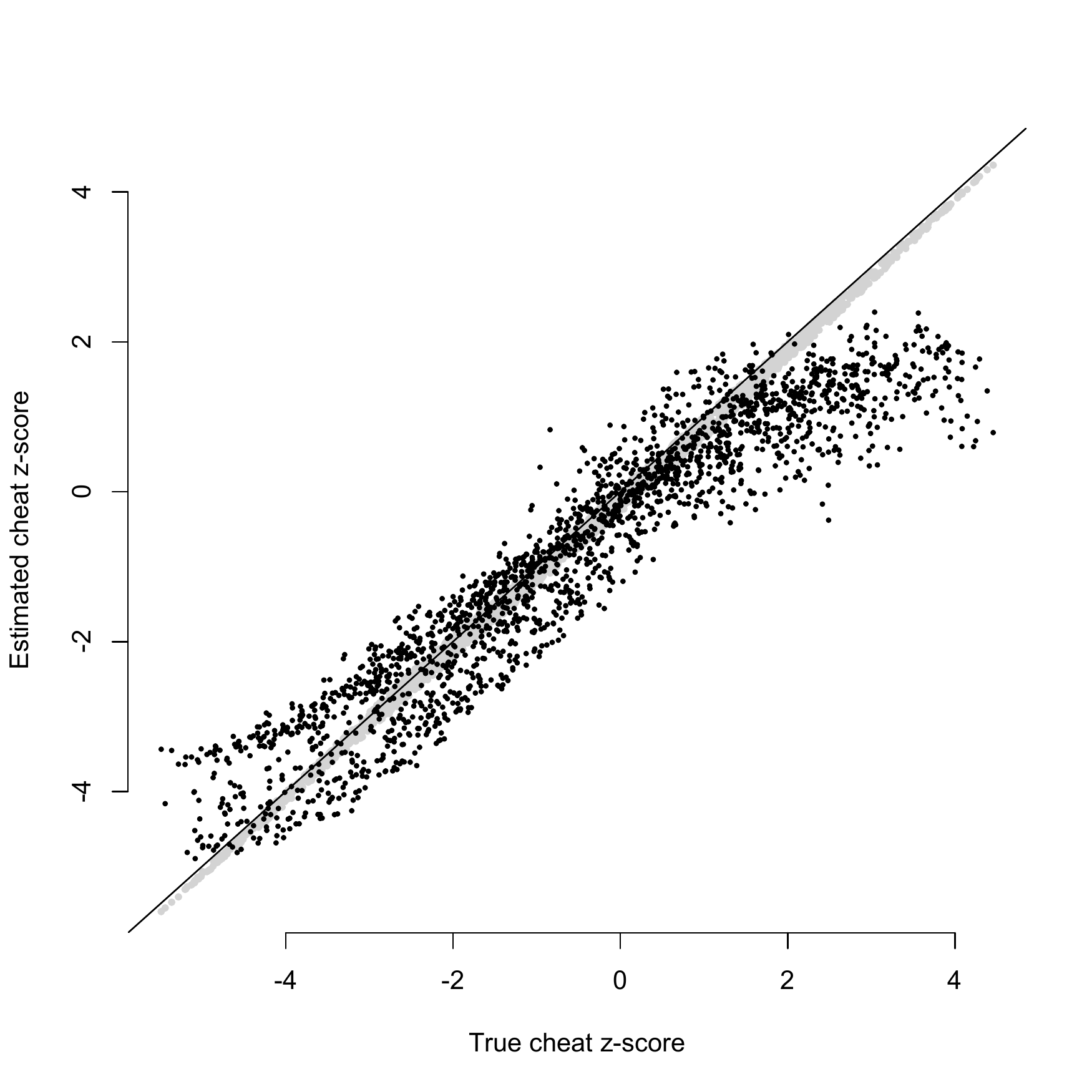}
\caption{Estimated probability of cheating versus the actual probability of cheat on the normal linear predictor scale for each of $n = 2000$ data points. Black solid dots show the modular prior approach, which loosely surrounds the diagonal, demonstrating the unresolved uncertainty in the partial identification approach.  The solid gray dots correspond to the \cite{Wang2013} model, which is correctly specified in this example; they hew more tightly to the diagonal.}\label{synthfig1}
\end{center}
\end{figure}

A Bayesian specification of \cite{Wang2013}, with vague conjugate priors for $\beta$ and $\gamma$ and a uniform prior on $(-1,1)$ for $\rho$, was fit using a Gibbs sampler algorithm with a Metropolis-Hastings update for $\rho$.  Our modular prior approach proceeds by fitting the BART model (with default priors as described in \cite{chipman2010bart}) to the observed data $(Y_i, \X_i)$ and constructing the posterior estimate of $\Pr(Z_i \mid \X_i)$ by dividing posterior samples of $\phi_{\X}$ by draws of $\theta_{\X}$ from (\ref{eq:theta-2}), with diagonal covariance $\sigma^2 \mbox{I}$ with $\sigma = 0.1$ and mean function set to match the true $\Pr(W = 1 \mid Z = 1)$ implied by (\ref{wang}). See the appendix for further computational details regarding model fitting. The results are depicted in Figure \ref{synthfig1}. As expected, the \cite{Wang2013} model, which achieves point-identification when it is correctly specified, yields much more accurate inference compared to our approach. Meanwhile, even with a correct surveillance model in this case, there persists a modicum of unresolved uncertainty, which reflects that in our model the estimand is only partially identified. Additionally, we see the impact of the BART prior pulling the estimated probabilities towards $1/2$ in regions near the edges where there are fewer data points.

\begin{figure} 
\begin{center}
\includegraphics[width=3.25in]{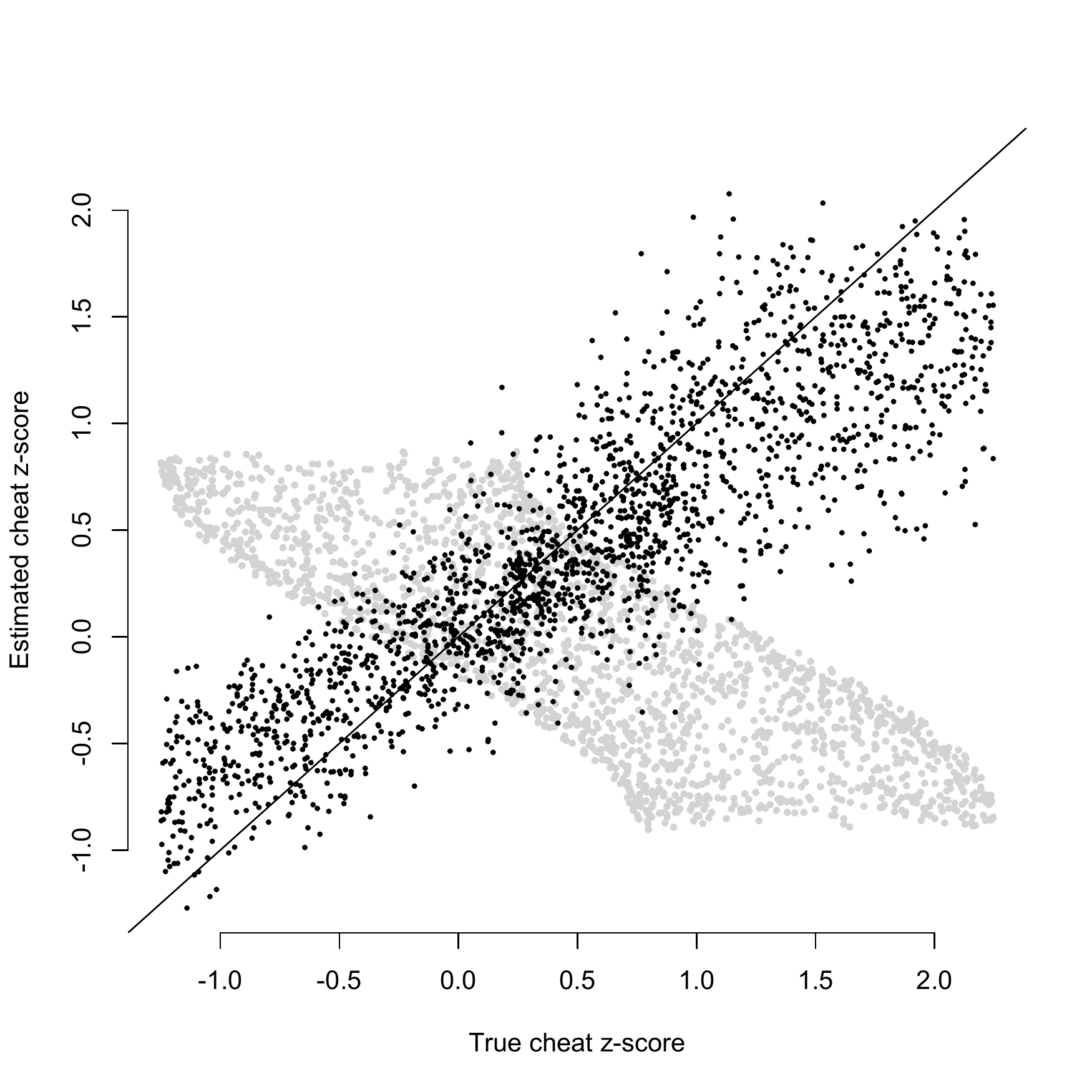}
\caption{Estimated probability of cheating versus the actual probability of cheat on the normal linear predictor scale for each of $n = 2000$ data points. Black solid dots show the modular prior approach, which loosely surrounds the diagonal, demonstrating the unresolved uncertainty in the partial identification approach.  The solid gray dots correspond to the \cite{Wang2013} model, which is incorrectly specified in this example; they grossly diverge from the diagonal.}\label{synthfig2}
\end{center}
\end{figure}

Our second demonstration deviates from the linear probit model by specifying $\mu(\X) = (0.5 + \sin{(\X + \pi)}^t\beta, \sin{(\X)}^t\gamma)$ for the same values of $\gamma$ and $\beta$.  The \cite{Wang2013} model is fit with the predictor variables unadjusted.  The modular prior approach is fit the same as before, using the correctly specified surveillance model. The results are depicted in Figure \ref{synthfig2}. As might be expected, under misspecification the \cite{Wang2013} model badly mis-estimates the true $\Pr(Z_i = 1 \mid \X_i)$.  Our approach, with good prior information, still does not achieve point identification, but manages to avoid the gross mis-fit of the \cite{Wang2013} model by successfully recovering the nonlinear identified component from the data. 
 
Naturally, if we had supplied invalid surveillance models, our approach may have been far off the mark in both cases. The point of this demonstration is merely that proceeding in a partially identified fashion is a more conservative course of action than choosing an implausible model on the grounds that --- {\em should it happen to be correct} --- it would deliver the desired point identification.

\section{Analysis of AAER data, 2004-2010}\label{empirical}
Our data are aggregated from three main sources.  First, the AAER response variable was obtained from the Center for Financial Reporting and Management (CFRM) at Berkeley's Haas School of Business.  Detailed information about the full data set can be found in \cite{dechow2011predicting}. 

Second, additional firm attributes are obtained from the CompuStat North America Annual Fundamentals database via the Wharton Research Data Service (WRDS).  These data are then merged with the AAERs using Global Company Key (GVKEY) by year.  Specifically, the covariates considered are:
\begin{itemize}\addtolength{\itemsep}{-0.5\baselineskip}
\item fiscal year,
\item cash,
\item net income,
\item capital investments,
\item SIC industry code,
\item qui tam dummy variable.
\end{itemize}
\noindent Cash, net income and capital investments are all recorded as a fraction of the firm's total assets.  Standard industrial classifications are given in terms of ten major divisions, denoted A-J by the Occupational Safety and Health Administration.  The {\em qui tam} dummy variable is derived from the SIC codes; it denotes if a firm is in an industry where persons responsible for revealing misconduct are eligible to receive some part of any award resulting from subsequent prosecution.  Similar to \cite{dyck2008corporate} and  \cite{jayaraman2010whistle}, our qui tam variable is set to one for firms with SIC code 381x, 283x, 37xx, 5122 or 80xx, which includes healthcare providers and pharmaceutical firms, and airplane, missile, and tank manufacturers.  It is reasonable to suppose that firms in these industries have a greater likelihood of any misconduct being exposed as a result of incentivized employees.  

Finally, a keyword search at the Financial Times web page (www.ft.com) was conducted on each company name and the number of search results was recorded by year.  This variable provides a crude measure of media exposure.  Although discrepancies between firm names as recorded in CompuStat and firm names as reported in Financial Times articles lead to measurement error of this variable, it still provides a  reasonable proxy for name recognition and cultural visibility. Most firms will never be mentioned in any news article; a few firms are routinely mentioned in the press. To adjust for the fact that web traffic has increased over that period, we normalize the search results count by the total number of hits across all companies in a given year.

We restrict our analysis to U.S. firms who had positive net income for the given year, considering the period between 2004 and 2010, for a total of $6,641$ unique firms and a total of $n=25,889$ total firm-year observations.
\subsection{Surveillance model}\label{selection}
The unidentified function $\theta(\X)$ can be thought of as a ``surveillance" probability; it encodes which attributes invite (discourage) SEC scrutiny, making cheaters more (less) likely to be caught.  Its effect is to inflate the probability of cheating, which is intuitive since the proportion of caught cheaters $\phi(x)$ represents an obvious lower bound on the proportion of actual cheaters.

Our surveillance model takes the form reported in expression (\ref{eq:theta-2}) with  $\Sigma_{\mathcal{X}} = \sigma^2 \mbox{I}$ and using a logit link.  We  scale and shift all variables to reside on the unit interval, taking shifted log transformations of the financial times search hits and net income.
%
%
We chose to place nonzero coefficients on the (fiscal) year of misconduct, media exposure (as measured by Financial Times search hits), income, cash, and a dummy for qui tam industries. We have specific reasons to expect that these variables are important determinants of the surveillance probability, allowing us to chose informative values for $\beta$.

\begin{figure}
\begin{center}
\includegraphics[width=3in]{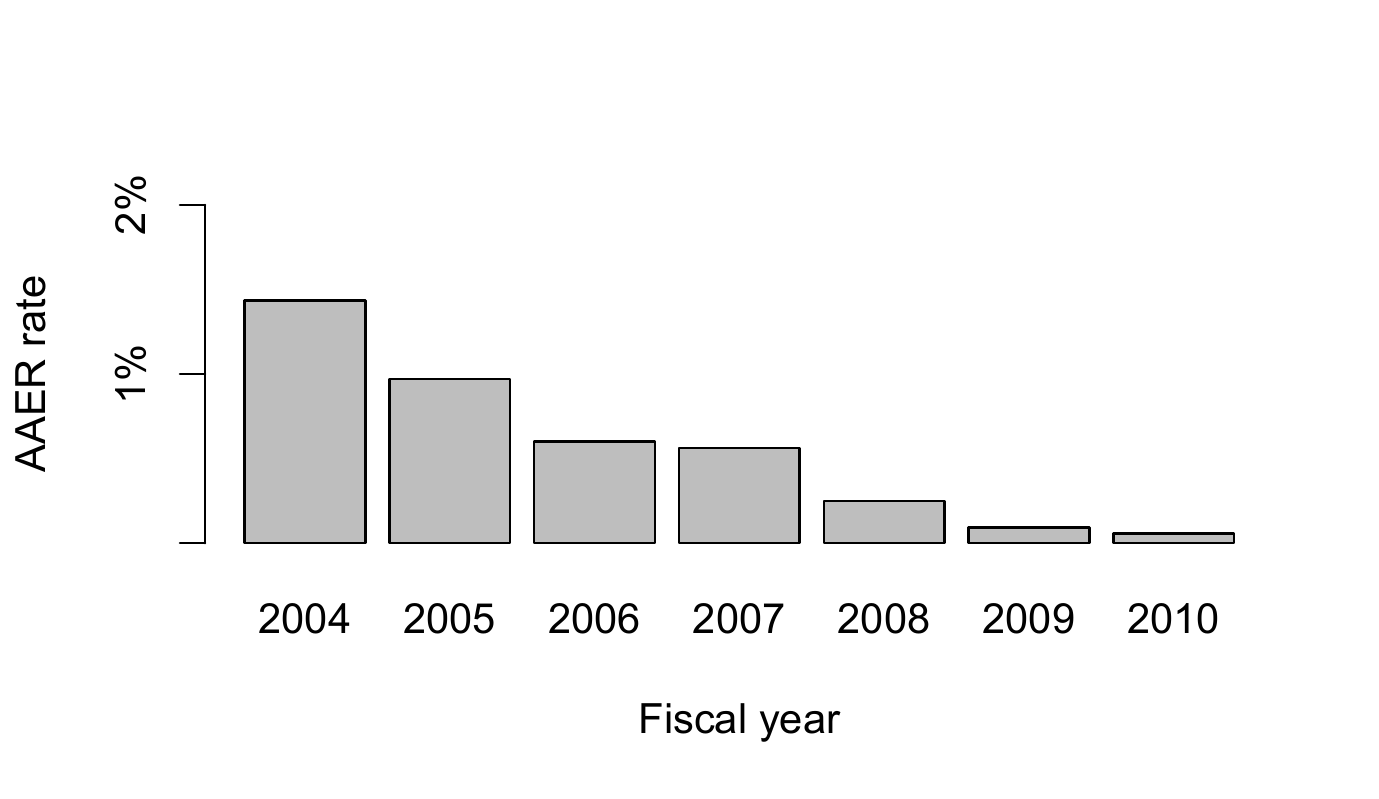}
\end{center}
\caption{AAERs appear more common in more distant years, but this is likely because they may be filed retroactively.}\label{aaers}
\end{figure}

First, note that the frequency of AAERs is substantially higher in earlier years; see Figure \ref{aaers}. AAERs may be filed retroactively, so the opportunity to discover and report misconduct in a given year increases over time.
Fitting a curve to the data in Figure \ref{aaers} suggests a coefficient $\beta_{fyear} = -2.5$.  Observe that this makes the posterior probability of cheating approximately constant across the years examined, which seems plausible.

Second, it is reasonable to assume that media attention naturally draws SEC scrutiny \citep{miller2006press}.  
The SEC has a vested interest in catching and making examples of any high-profile cheaters.
We set $\beta_{FThits}=2$, implying approximately a six-fold difference in the probability of misconduct being discovered between a company with no media exposure and the company with the highest media exposure. Similarly, we set $\beta_{quitam}=1$, implying approximately a two-fold increase in misconduct being discovered for companies in qui tam industries where employees are incentivized to report misconduct.

These observations constitute the subjective information contained in our first observation model (A).  
To determine the intercept term, consider the following argument: AAERs are quite rare, with an aggregate incidence in our sample of only 0.5\%.  Potentially this is because very few firms exhibit actionable misconduct, but more likely it is because the SEC has limited resources to identify and pursue violations.  Accordingly, one sensible calibration method would be fix the mean probability of discovery across all firms.  Fixing this quantity and the other elements of $\beta$, we may then solve for the intercept $\beta_0$. In the case of model A, fixing the average discovery rate to 30\% gives $\beta_0 = 0$.

After obtaining posterior samples under model A, we observe that cash appears to have a negative impact on the probability of cheating. In contrast, net income appears to have a positive impact on the probability of cheating. We might surmise that these trends are due to unadjusted surveillance probabilities.  For example, one could argue that having large amounts of cash on hand (relative to total assets) provides a measure of ``wiggle room" that makes certain kinds of misconduct harder to discover.  Likewise, firms with high income are more likely to draw SEC attention than firms with smaller income streams.  To compare our results under these narratives, we specify a second surveillance model (B), with $\beta_{cash} = -1.5$ and $\beta_{income} = 2.5$.  Setting the intercept for model B to match the 30\% discovery rate of model A gives $\beta_0 = -0.85$.  Surveillance models  A and B are shown side-by-side in Table \ref{surveillance}.

\begin{table}
\begin{center}
\begin{tabular}{c|cccccc}
&Intercept&Fiscal year &FT.com hits&cash&net income&qui tam\\
\hline 
$\beta_A$ & $\;\;\;0$& $-2.5$&$2$&$0$& $0$& $1$\\ 
$\beta_B$ &$-0.85$&$-2.5$& $2$&$-1.50$&$2.5$&$ 1$\\
\end{tabular}\caption{Regression coefficients defining surveillance models A and B.  They differ in their intercept terms and their cash and net income coefficients.  The intercepts have been adjusted to obtain an average misconduct discovery rate of 30\%}\label{surveillance}
\end{center}
\end{table}

\subsubsection{Upper bounding cheating}\label{cparm}
Finally, we introduce an additional parameter $c$ that allows us to interject prior information concerning an upper bound on the probability of cheating.  Recall that the partial identification in this application is driven by the inequality $\mbox{Pr}(Z = 1 \mid \X) = \frac{\phi(\X)}{\theta(\X)} \leq 1$, which implies $\phi(\X) \leq \theta(\X)$; the left hand side of this latter inequality is identified by the data.  
Extremely high probabilities of cheating are implausible, motivating us to consider alternative truncations: $\mbox{Pr}(Z = 1 \mid \X) = \frac{\phi(\X)}{\theta(\X)} \leq c$, which implies $\phi(\X)/c \leq \theta(\X)$ for $c \geq \phi(\X)$ for all $\X$.  Because $\phi(\X)$ is identified, the data may contradict any particular value of $c$ by violating $c \geq \phi(\X)$ for some $\X$.  That is, $\eta := c$ and $\Omega\{\phi(\X)\} = \{ (c, \theta(\X)) \mid \phi(\X)/c \leq \theta(\X) \leq 1\}$. Note that $\eta := c$ is ``global" in that it implies inequalities across all values of $\X$, whereas $\theta_{\X}$ can take distinct values at distinct $\X$ locations.  

We specify $\pi(c \mid \phi(\X)) \propto \mbox{Beta}\left(10 c_0, 10(1-c_0) \right)\mathbb{1}\{c \geq\sup_{\X} \phi(\X)\}$.  The prior mean $c_0$ captures prior beliefs about the upper bound on $\mbox{Pr}(Z = 1 \mid \X)$, the probability of cheating for any firm.  This permits us to specify $\pi(\theta_{\X} \mid c, \phi_{X}) \propto \N(h(\X)\beta, \sigma^2)\mathbb{1}\{\theta_{\X} \geq \phi_{\X}/c\}$; joint samples of $(c, \theta_{\X})$ are thereby guaranteed to lie in $\Omega(\phi_{\X})$.

Our surveillance models allow us to include prior information in the form of
subject matter knowledge about the impact of various covariates. 
We are also able to include additional subjective prior information about $\sum_i \mbox{Pr}(W_i = 1 \mid Z_i = 1, \X_i)$ --- the overall average probability of a cheating firm getting caught --- via the intercept terms, and $\mbox{max}_i \mbox{Pr}(Z_i = 1 \mid \X_i)$ --- an upper probability on any firm cheating --- via the prior on $c$. 

Further computational details are included in the Appendix.

\subsection{Results}
We conduct a sensitivity analysis by varying the parameters $\sigma$ and $c$ for both model coefficients $\beta_A$ and $\beta_B$ above.  Specifically, we consider two settings of each ($\sigma \in \lbrace 0.25,0.5 \rbrace$ and $c \in \lbrace 0.4,0.8 \rbrace$) for a total of eight candidate models.  We study the impact these choices have on both $\mbox{Pr}(Z=1 \mid \X)$ as a function of individual predictor variables, and also on the overall misconduct prevalence $\alpha$.

\begin{figure}
\begin{center}
\includegraphics[width=\textwidth]{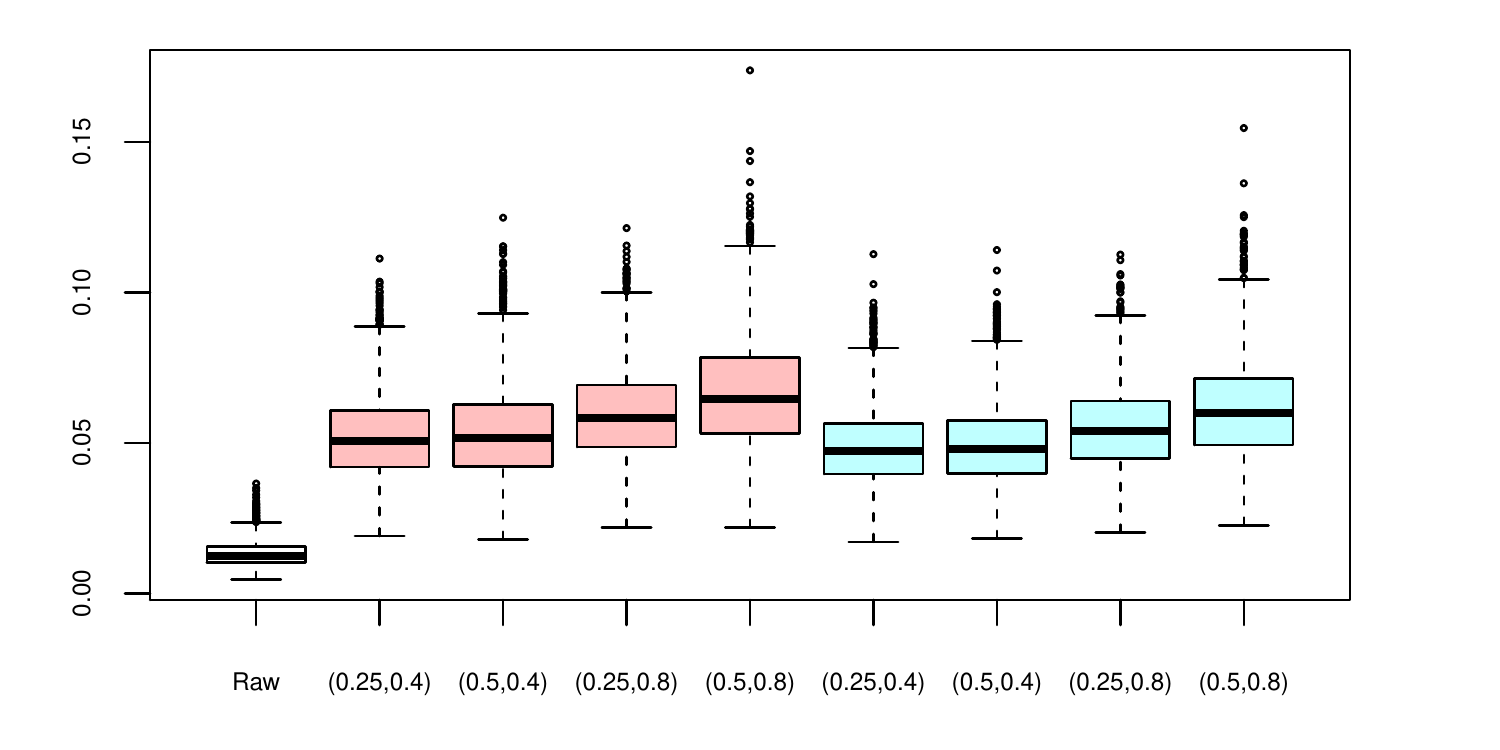}\\
\includegraphics[width=\textwidth]{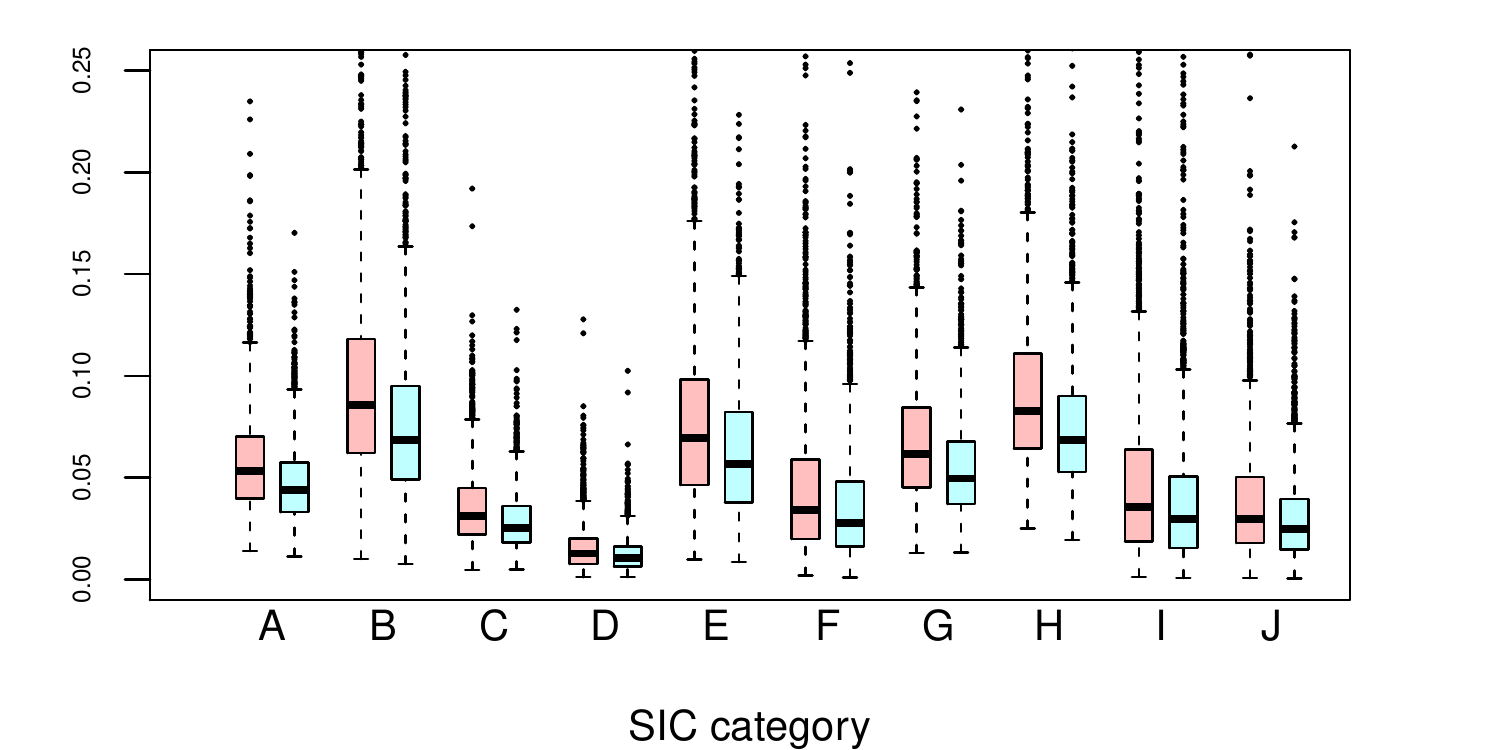}
\end{center}
\caption{Top panel: posterior cheating prevalence, white corresponding to the raw (unadjusted) prevalence, pink to prior A and blue to prior B. The four boxplots within each prior correspond to the following combinations for $c$ and $\sigma$ (from left to right): $(\sigma=0.25,c = 0.4),\;(\sigma=0.25, c = 0.8 ),\;(\sigma=0.5, c=0.4),\;(\sigma=0.5, c= 0.8)$. Bottom panel: Posterior cheating prevalence in companies within each SIC code, pink corresponding to prior A and blue to prior B for $(\sigma=0.25,c=0.4)$.  }\label{SICboxplot}
\end{figure}

Figure \ref{SICboxplot} shows the posterior distribution of the adjusted cheating prevalence under the different models. As can be seen in the top panel, increasing $\sigma$ or $c$ alone has little effect on the overall adjusted cheating prevalence. These two parameters control different aspects of the surveillance uncertainty: a high $c$ implies that any probability of cheating is plausible, whereas a high $\sigma$ allows large deviations from the specified surveillance model logistic term.  Note that with the range of $c$ and $\sigma$ chosen, neither surveillance model yields overall prevalence of misconduct greater than 15\%.

The bottom panel of Figure \ref{SICboxplot} shows the posterior prevalence for each different SIC code under the two priors, fixing $(\sigma=0.25, c = 0.4)$. Under both priors, SIC category D, representing ``electricity, gas, steam and air conditioning supply'', shows much lower cheating prevalence than categories (B,E,H), which correspond to ``mining and quarrying'', ``water supply, sewerage, waste management and remediation activities'', and ``transportation and storage'' respectively. 

Because BART is nonlinear, summary plots of the impact of individual covariates are challenging to visualize, even if the surveillance model is relatively simple, such as our linear logistic specification.  It is instructive, therefore, to examine the implied probability of cheating as one varies individual covariates, for a given firm.  That is, how does $\mbox{Pr}(Z=1 \mid \X)$ change as a function of $x_j$ while holding $x_{-j}$ fixed?

\begin{figure}
\begin{center}
\includegraphics{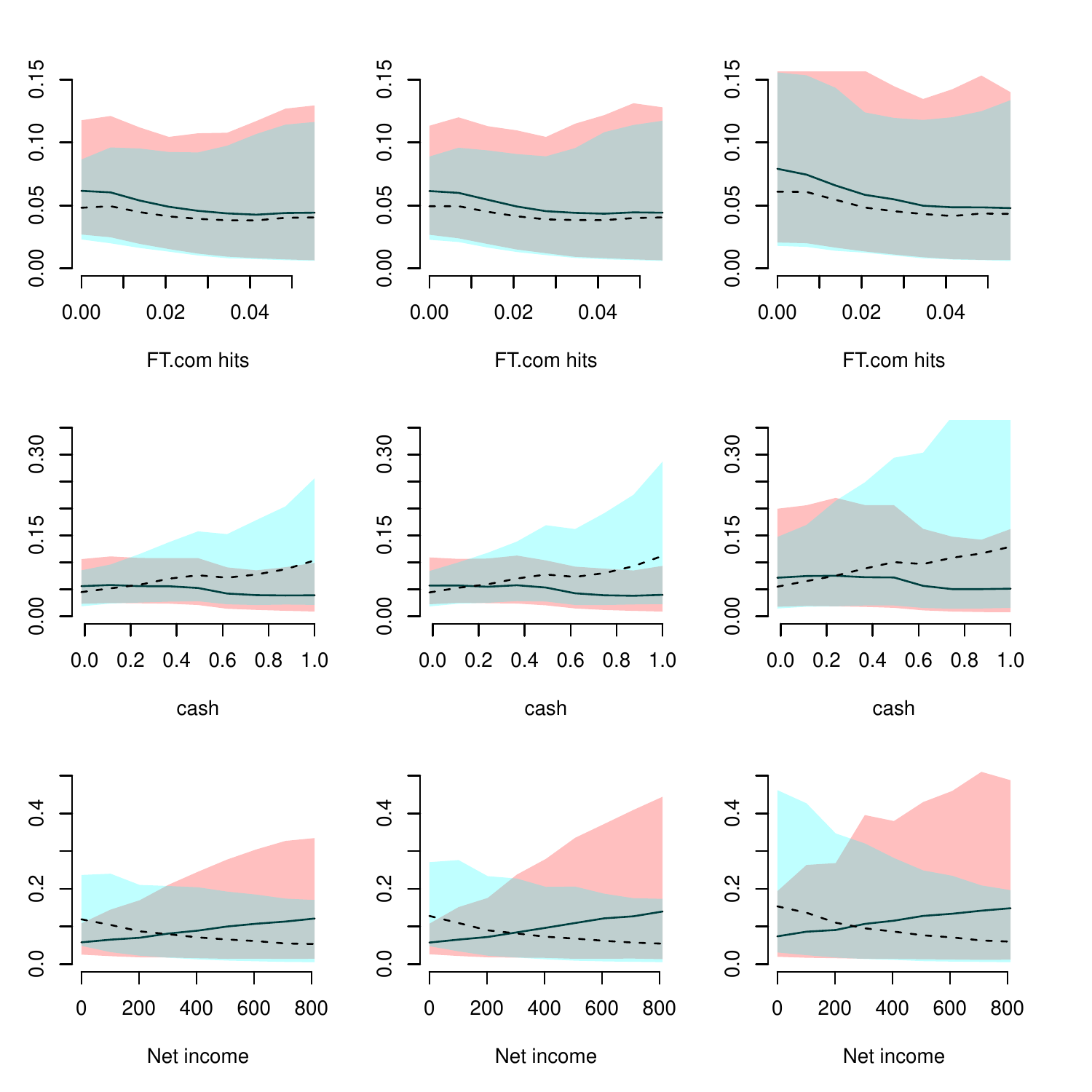}
\end{center}
\caption{$\mbox{Pr}(Z=1 \mid \X)$ varying $x_j$ while holding $x_{-j}$ fixed (at the values of ConAgra Foods) under two surveillance models A (pink) and B (blue). Lines depict the median and shaded areas depict 90\% credibility intervals. Each row represents a different covariate; the three columns correspond to the following $(c, \sigma)$ combinations (from left to right): $(c=0.4,\sigma=0.25),\;(c=0.8,\sigma=0.25),\;(c=0.8,\sigma=0.5)$. }\label{VaryPriorVarySC}
\end{figure}

To demonstrate this approach, we focus on a specific firm, ConAgra Foods of Omaha, Nebraska (simply because it yields illustrative plots).  Figure \ref{VaryPriorVarySC} shows $\mbox{Pr}(Z=1 \mid \X)$, varying media attention, cash, and net income under priors A and B and for various combinations of $c$ and $\sigma$. As expected, the surveillance model coefficients on cash and net income reverse the associated slope of the probability of cheating.  High values of both $\sigma$ and $c$ results in posterior credible intervals of up to 40\% probability of cheating for some values of net income. 

We have reported here only a small number of the possible variations one would presumably want to investigate; we make no claims that models A and B are ideal or even necessarily good or realistic models.  Rather, our sensitivity analysis demonstrates a range of possible comparisons that one might undertake when investigating how various assumptions interact with the data via the identified portion of the model.  

An important upshot of our analysis is that the surveillance model intercept terms---which govern the  average probability of misconduct discovery (getting caught) across firm---and the parameter $c$--- which defines the upper bound $\mbox{Pr}(Z_i =1 \mid \X_i) \leq c$ for all $i$ --- play dominant roles in determining the inferred overall prevalence of misconduct.  For our choices of 30\% misconduct discovery probability and $c = 0.4$ or $c=0.8$, we find that no more than 15\% of firms engage in accounting misconduct.  

Incidentally, this result is consistent with that of \cite{Dyck2013}, who put the prevalence at between 4.74\% to 15\%, based on a clever natural experiment resulting from the dissolution of the large accounting firm Arthur Andersen and the subsequent re-audit of its clients following the Enron scandal. Unavailability of their exact data, as well as the unavailability of the  data of \cite{Wang2013} at the time of writing, means that we cannot compare their precise estimates with those from our model.  However, our partial identification analysis suggests that any similar analysis is likely to yield similar conclusions in the matter of overall prevalence.  After all, there is only so much information in the available data, with the rest coming from auxiliary assumptions about the surveillance probability, whether those assumptions are explicit, as in our model, implicit, as in the joint likelihood assumed by \cite{Wang2013}, or based on supplementary evidence, as in \cite{Dyck2013}. To the extent that these various approach supply similar assumptions, they will yield similar conclusions.  Our approach, by layering such assumptions over the data {\em ex post}, permits systematic sensitivity analysis rather than one-off comparisons of published studies whose authors are committed to one particular approach.

\section{Discussion}
Working directly with modular priors in partially identified settings has several advantages. First, it allows identified parameters to be modeled flexibly, permitting the data to be maximally informative, while simultaneously allowing the analyst to specify informative priors for the underidentified components of the model.  It may appear that this tactic stands in contrast to the approach of \cite{gustafson2010bayesian} (for example) which advocates working with a scientific model directly in the $\tau$ parametrization. However, nothing in our approach precludes the use of such subject-specific information.  Rather, we argue that typical prior specifications for $\tau$ do not allow separately modulating the prior informativeness on the identified and unidentified components; by working directly in the $(\phi, \theta)$ representation,  we achieve precisely this sort differential informativeness.  Nonetheless, one should always be mindful of the implied prior on $\tau$.  Specifically, analysts can use intuitions regarding $\tau$ as a tool for vetting priors over $(\phi, \theta)$, by checking (via simulation) that they are consistent with available knowledge in the $\tau$ representation.  In many applications, such as the one studied in this paper, the $(\phi, \theta)$ representation is itself readily interpretable (in this case, the ``cheating probability" and the ``surveillance probability", respectively).

Second, when an interpretable parameterization of the modular parameters is available (as it is in our application), the modular prior approach facilitates efficient sensitivity analyses. Sensitivity analysis is good practice generally, and vital when the data are completely uninformative about certain aspects of the model. Being able to conduct such analyses without refitting the entire model can be a tremendous practical advantage, particularly when fitting sophisticated nonlinear regression models to the identified component.

Finally, while we have not emphasized it in our applications, our use of modular priors allows us to not only conduct sensitivity analyses, but also perform model checking on the unidentified component. By computing ``residuals" for our model of $\theta(\X)$ based on the support function $\Omega\{\phi(\X)\}$ we can learn about areas in the covariate space where our prior model is contradicted by the data---that is, regions where the model places significant mass outside $\Omega\{\phi(\X)\}$. This is a unique opportunity in partially identified models that, to our knowledge, has not been exploited in the existing literature. Performing these model checks in our applications did not raise any red flags, but they would indicate potential problems in situations such as our example in section \ref{sec:ex}.

We expect these methods to be useful in other contexts with partially observed data, including traditional missing data settings (nonignorable unit or item survey nonresponse, dropout, or measurement error) \citep{rubin1976inference, little1987statistical, Daniels2008}, causal inference with observational data or imperfectly randomized study designs \citep{imbens1997bayesian, barnard2003principal, greenland2005multiple, McCandless2007, McCandless2012}, and ecological inference \citep{duncan1953alternative, shively1969ecological,AchenShively199505, king2013solution}, to name just a few. Our approach allows practitioners to consider much more sophisticated regression modeling for identified parameters while taking a principled approach to the lack of identification, and promotes the routine use of sensitivity analysis and model checking due to the ease with which they are implemented. For an illustration of our approach applied to an imperfectly randomized study of flu vaccine efficacy, see the supplementary material.

\singlespace 
\bibliographystyle{plainnat}
\bibliography{bib/isolating_causalB}
\doublespace
\appendix 
\section{Bayesian additive regression trees}
In our application, the nonlinear function $\phi(\X)$ was modeled using the Bayesian Additive Regression Trees (BART) approach of \cite{chipman2010bart}. Though nothing in our method is specific to this choice, the BART model has several properties that make it a sensible choice.  BART is more flexible than classical parametric regression models, such as linear logistic or probit regression \citep{cox1958regression}, but, unlike alternative nonparametric Bayesian regression models, BART is able to detect interactions and discontinuities and is invariant to monotone transformations of the covariates. For clarity, in this appendix we keep to the original notation used in \cite{chipman2010bart}; several of the BART prior parameter names are used in the main text where they refer to different entities.

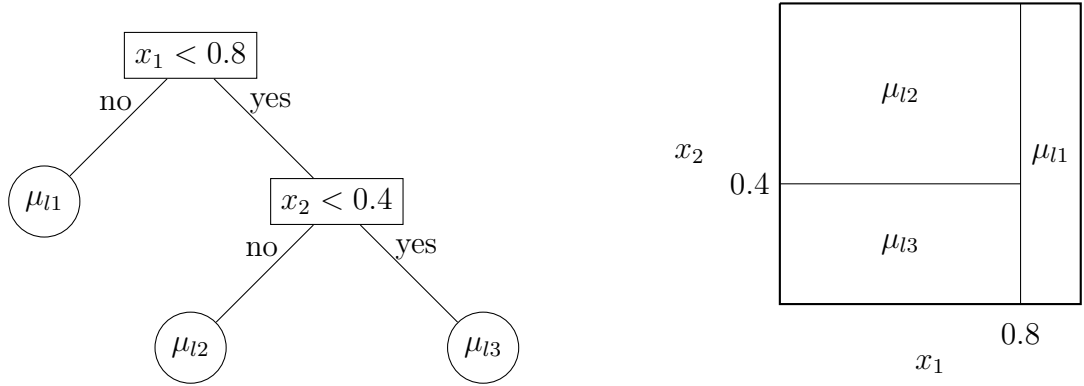
\begin{figure}
\begin{center}
\begin{tikzpicture}[
  scale=1.1,
    node/.style={%
      draw,
      rectangle,
    },
    node2/.style={%
      draw,
      circle,
    },
  ]

    \node [node] (A) {$x_1<0.8$};
    \path (A) ++(-135:\nodeDist) node [node2] (B) {$\mu_{l1}$};
    \path (A) ++(-45:\nodeDist) node [node] (C) {$x_2<0.4$};
    \path (C) ++(-135:\nodeDist) node [node2] (D) {$\mu_{l2}$};
    \path (C) ++(-45:\nodeDist) node [node2] (E) {$\mu_{l3}$};

    \draw (A) -- (B) node [left,pos=0.25] {no}(A);
    \draw (A) -- (C) node [right,pos=0.25] {yes}(A);
    \draw (C) -- (D) node [left,pos=0.25] {no}(A);
    \draw (C) -- (E) node [right,pos=0.25] {yes}(A);
\end{tikzpicture}
\hspace{0.1\linewidth}
\begin{tikzpicture}[scale=4]
\draw [thick, -] (0,1) -- (0,0) -- (1,0) -- (1,1)--(0,1);
\draw [thin, -] (0.8, 1) -- (0.8, 0);
\draw [thin, -] (0.0, 0.4) -- (0.8, 0.4);
\node at (-0.1,0.4) {0.4};
\node at (0.8,-0.1) {0.8};
\node at (0.5,-0.2) {$x_1$};
\node at (-0.3,0.5) {$x_2$};
\node at (0.9,0.5) {$\mu_{l1}$};
\node at (0.4,0.7) {$\mu_{l2}$};
\node at (0.4,0.2) {$\mu_{l3}$};
\end{tikzpicture}
\end{center}
\caption{(Left) An example binary tree, with internal nodes labelled by their splitting rules and terminal nodes labelled with the corresponding parameters $\mu_{lb}$ (Right) The corresponding partition of the sample space and the step function.}
\label{fig:treestep}
\end{figure}

The BART prior represents an unknown function $f(\X)$ as a sum of many piecewise constant binary regression trees.
%
Each tree $T_l,\;1\leq l\leq L$, consists of a set of internal decision nodes which define a partition of the covariate space (say $\mathcal{A}_1,\dots,\mathcal{A}_{B(l)}$), as well as a set of terminal nodes or leaves --- one corresponding to each element of the partition. Each subset of the partition $\mathcal{A}_b$ is associated with a parameter value $\mu_{lb}$, defining a piecewise constant function: $g_l(x) = \mu_{lb}\ \text{if}\ x\in \mathcal{A}_b.$ This regression tree function representation is depicted in Figure \ref{fig:treestep}. 

Individual regression trees are then additively combined into a single regression function: $f(\X)=\sum_{l=1}^L g_l(\X).$ The representation of $f(\X)$ through the sum of a set of regression trees is generally non-unique; in our applications, this redundancy is unproblematic.

Each of the functions $g_l$ are constrained by their prior to be ``weak learners"; that is, the prior strongly favors small trees and leaf parameters that are near zero. Each tree independently follows the prior described by \cite{chipman1998bayesian}, where the probability that a node at depth $d$ splits (is not terminal) is given by 
$\alpha (1+d)^{-\beta},\;\;\alpha\in (0,1),\;\beta\in [0,\infty).$

A variable to split on, and a cut-point to split at, are then selected uniformly at random from the available splitting rules.  Large, deep trees are given extremely low prior probability by taking $\alpha=0.95$ and $\beta=2$ as in \cite{chipman2010bart}. 
The leaf parameters are assigned independent priors
$\mu_{lb}\sim N(0,\sigma^2_\mu)\;\;\textrm{ where }\sigma_\mu=3/(k\sqrt{L})$. The default  value, $k=2$, shrinks $g_l(x)$ strongly toward zero. The induced prior for $f(x)$ is centered at zero and puts approximately 95\% of the prior mass within $\pm 3$ pointwise. Larger values of $k$ imply increasing degrees of shrinkage.

Finally, the probability function of interest is modeled as $\Phi\left(f(\X)+c\right)$, where $c$ is an offset parameter and $\Phi(\cdot)$ is the standard normal cumulative distribution function. Complete details of the BART prior and its implementation are given by \cite{chipman2010bart}.

\subsection{Computation}
Posterior inferences are obtained by sampling at a fixed grid of design points $\X^*_1, \dots, \X^*_J$.  To reduce notational clutter, we suppress dependence on $\X_j^*$; for example, $\theta$ should be read as referring to a specific $\theta(\X_j^*)$.

Operationally, posterior samples are obtained according to the following recipe.

\begin{enumerate}
\item Fit the BART model to the observed pairs $(Y_i, \X_i)$, for $i = 1, \dots, n$.  Convenient {\tt R} implementations facilitate this step readily, for example {\tt BayesTree} or {\tt dbarts}. This gives a collection of posterior samples $\phi^1, \dots, \phi^k$.
\item For each posterior sample $\phi^k$, draw $\theta^k$ by first drawing $\eta^k$ from $\pi(\eta^k \mid \phi^k)\mathbb{1}\{\eta^k \in \Omega[\phi^k]\}$ and then drawing $\theta^k$ from $\pi(\theta^k \mid \phi^k, \eta^k)\mathbb{1}\{\theta^k,\eta^k \in \Omega[\phi^k]\}$.
\end{enumerate}
See \cite{Chan2014} for a similar computational approach. Because sampling from the posterior of $\phi$ operates independently from sampling the partially identified parameter $\theta$ --- as shown in (\ref{composition}) in the main text --- sensitivity analysis can be conducted without ever needing to refit the model,  simply by repeating step 2 for various choices of  $\pi(\eta, \theta \mid \phi)$.

In our empirical application, we use the Gaussian process prior (\ref{eq:theta-2}) with diagonal covariance matrix, $\eta := c$ is given a Beta prior as described in Section \ref{cparm}, and we use a standard normal probit link for $F^{-1}$.  For these choices, step 2 above becomes:
\begin{enumerate}[label=(\roman*)]
\item Draw $c$ from its truncated Beta distribution with lower truncation point given by $\max\left\{\phi(\X^*_1),\ldots,\phi(\X^*_J)\right\}$.
\item  Draw $F^{-1}\{\theta_{\X}\}$ from independent truncated normal distributions at each design point, with lower bound $F^{-1}\{\phi_{\X}/c\}$.  
\end{enumerate} 

\clearpage 
\section{Supplementary Application: Randomized encouragement study of flu vaccine }\label{flu}

Our data come from a randomized encouragement study of the effectiveness of an influenza vaccine conducted at an academic primary care practice affiliated with a large teaching hospital in an urban setting \citep{McDonald1992}. Physicians were randomly assigned to treatment or control groups, with treated physicians receiving a computerized reminder about the vaccine when a patient with a scheduled visit also met the U.S. Public Health Service Criteria for receiving a flu vaccine. Patients were eligible for the vaccine if they were 65 or older, or if they had certain severe chronic diseases including chronic obstructive pulmonary disease (COPD).
The outcome of interest is subsequent hospitalization for flu-related illness, and the covariates include the patient's age and COPD status. \cite{Hirano2000} conduct a Bayesian analysis of this problem, placing a default joint prior over all the identified and unidentified parameters. In this application we perform an alternative selectively informative analysis with modular priors. 
We will use the terms ``reminder'', ``instrument'' and ``encouragement'' interchangably here, prefering ``reminder'' in the context of the specific applied problem.
 
\subsection{Causal Inference in a Binary Instrumental Variables Model}

Let $Z_i, T_i$, and $Y_i$ be the observed values of the treatment assignment (or encouragement, or instrument), treatment received, and outcome for observation $i$, respectively.  Let $X_i$ be a vector of baseline covariates. In the flu data, $Z_i=1$ if encouragement (the reminder) was administered, $T_i=1$ if the vaccine was administered, and $Y_i=1$ if the patient was \emph{not} hospitalized. As in \cite{Angrist1996} and \cite{Hirano2000} define the \emph{potential outcomes} $T_i(z)$ and $ Y_i(t,z)$ as the treatment
assignment that would have been observed had we fixed $Z_i=z$ and the outcome we would have observed fixing $T_i=t, Z_i=z$, respectively. We will suppress indexing by $i$ for population quantities. We make the stable unit treatment value assumption (SUTVA) \citep{Angrist1996}, i.e. that there is no interference between units and no alternative versions of treatment. \footnote{\cite{Hirano2000} point out that in an infectious disease application with individuals as the unit of randomization this assumption is unlikely to hold exactly, but constitutes a reasonable first order approximation.} Mathematically, SUTVA implies that $T_i(\mathbf{z})=T_i(z_i)$ and $Y_i(\mathbf{t}, \mathbf{z})=Y_i(t_i, z_i)$, where $\mathbf{z}, \mathbf{t}$ are vectors of instrument/encouragement assignment and treatments received (respectively) for the entire sample.

In this application the complete vector of binary response variables is 
\begin{equation}
U_i=(T_i(0), T_i(1), Y_i(0,0), Y_i(0,1), Y_i(1,0), Y_i(1,1))',
\end{equation}
 the collection of treatments and outcomes under all possible experimental conditions. However, we observe only $T_i\equiv T_i(Z_i)$ and $Y_i\equiv Y_i(T_i(Z_i), Z_i)$, where $Z_i$ is randomly assigned at the study outset. 
%
%
We can define compliance types or principal strata $S$ based on the possible values for $(T(1), T(0))$ -- see Table \ref{tab:comptype}. 
\begin{table}
\begin{center}
    \begin{tabular}{cc|ll}
        $T(1)$ & $T(0)$ & $S$ (Compliance Type) & ~    \\ \hline
        0     & 0     & Never Taker     & (n) \\ 
        1     & 1     & Always Taker    & (a) \\ 
        1     & 0     & Complier        & (c) \\ 
        0     & 1     & Defier          & (d) \\
    \end{tabular}
  \caption{Compliance types, or principal strata, defined by response patterns across levels of $Z$.}
  \label{tab:comptype}
  \end{center}
\end{table}
We will assume that the encouragement $Z$ is ignorable or randomly assigned (as it is in the flu data).
We also assume that there are no defiers (``monotonicity of compliance"), i.e., $ \Pr(S=d\mid \X)=0$. In the current setting a defier would be a patient who would \emph{only} receive the vaccine if the reminder was \emph{not} administered, which is quite unlikely.

Let $\pi_s(x) = \Pr(S=s\mid \X)$  and 
$\gamma^{tz}_s(x) = \Pr( Y(t,z)=1\mid S=s, \X)$ denote the principal strata and potential outcome probability models. Under the above assumptions the potential outcome and principal strata distributions are linked to the joint distribution of observables by the following equations.
\begin{align}
\Pr(T=1\mid Z=0, \X) &= \pi_a(\X)\label{eq:mon-1},\\
\Pr(T=1\mid Z=1, \X) &= \pi_c(\X) + \pi_a(\X)\label{eq:mon-2},\\
\Pr(Y=1\mid T=0, Z=0, \X) &= [\pi_c(\X)\gamma^{00}_c(\X) + \pi_n(\X)\gamma^{00}_n(\X)]/[\pi_c(\X) + \pi_n(\X)]\label{eq:mon-3},\\
\Pr(Y=1\mid T=1, Z=1, \X) &= [\pi_c(\X)\gamma^{11}_c(\X) + \pi_a(\X)\gamma^{11}_a(\X)]/[\pi_c(\X) + \pi_a(\X)]\label{eq:mon-4},\\
\Pr(Y=1\mid T=0, Z=1, \X) &= \gamma^{01}_n(\X),\\
\Pr(Y=1\mid T=1, Z=0, \X) &= \gamma^{10}_a(\X)\label{eq:mon-last}.
\end{align}
\cite{Richardson2010} derive this mixture relationship between principal strata/potential outcome distributions and the observable joint in detail under a range of assumptions about the latent variables (see also \cite{Imbens1997a} and \cite{Richardson2011}).

The probabilities on the left hand side completely determine the joint distribution of observables ($Y$ and $T$ given $Z$) and hence constitute a basis for $\phi$. Clearly $\pi$, $\gamma^{01}_n$ and $\gamma^{10}_a$ are one-to-one maps from $\phi$, and are point identified as well. However, $\gamma^{00}_c,\ \gamma^{00}_n,\ \gamma^{11}_c,$ and $ \gamma^{11}_a$ are not. For example, at any point in covariate space $\gamma^{11}_c(x)$ and $\gamma^{11}_a(x)$ must lie on the line segment given by the intersection of the line in equation \eqref{eq:mon-4} and $[0,1]\times [0,1]$, but the data are wholly uninformative about their position on that line segment. In particular, the range of $\gamma^{11}_a(x)$ is

\begin{equation}
\left(
	\min\left(
		0, \frac{\pi_a(x)+\pi_c(x)}{\pi_a(x)}p_{1\mid 11}(x) - \frac{\pi_c(x)}{\pi_a(x)}
	\right),\,
	\max\left(
		1, \frac{\pi_a(x)+\pi_c(x)}{\pi_a(x)}p_{1\mid 11}(x)
	\right)
\right)\label{eq:gbound}
\end{equation}
where $p_{1\mid 11}(x) = \Pr(Y=1\mid T=1, Z=1,\X)$. The situation is exactly the same for $\gamma^{00}_c(x)$ and $\gamma^{00}_n(x)$, with the line given by \eqref{eq:mon-3} and bounds similar to \eqref{eq:gbound} (large sample bounds on causal effects within principal strata were introduced by  \cite{zhang2003estimation}; see \cite{Richardson2011} for bounds under monotonicity and other assumptions, as well as further references).

Point identification for the remaining parameters is often obtained through further assumptions on the distribution of the potential outcomes. The \emph{stochastic exclusion restriction} for compliance type $s$ posits that there is no direct effect of $Z$ on $Y$; that is, $\gamma^{i0}_s = \gamma^{i1}_s$ for $i=0,1$ \citep{Hirano2000}. From \eqref{eq:mon-1}-\eqref{eq:mon-last} it is clear that this achieves point identification for the entire distribution of the potential outcomes.

\cite{Hirano2000} anticipated failures of the exclusion restrictions, noting that a reminder about flu shots might also prompt physicians to counsel patients about other preventative measures to avoid contracting the flu. They propose relaxing the exclusion restrictions and resolving the resulting nonidentifiability through prior probability modeling. In this sense our approach and theirs are philosophically similar, but our framework affords us significant flexibility in choosing the model for $\phi$ while transparently embedding prior information about $\theta$, which is difficult to do in their modeling setup. In our subsequent analysis we will avoid imposing any exclusion restrictions.

\subsection{Model Specification}

Our model for observables decomposes into a treatment model and an outcome given treatment model, i.e., $P(Y,T\mid Z, \X) = P(T\mid Z, \X)P(Y \mid T, Z, \X)$. We assign each of these a probit BART prior. For the treatment model, randomization of treatment assignment and monotonicity of compliance imply that $\Pr(T=1\mid Z=1, \X) \geq \Pr(T=1\mid Z=0, \X)$ for all $\X$ (see equations \eqref{eq:mon-1} and \eqref{eq:mon-2}). We enforce this constraint by rejecting samples from BART where this condition fails at any point in covariate space. More elaborate approaches to enforcing monotonicity are possible but beyond the scope of the present paper. 

\subsubsection{Partially identified parameters}

Recall that we have two pairs of unidentified parameters,  $(\gamma^{00}_c, \gamma^{00}_n)$ and $(\gamma^{11}_c,\gamma^{11}_a)$. Fixing one element of each pair would solve the identification problem, so we can take $\theta = (\gamma^{00}_n,\gamma^{11}_a)$ without loss of generality. 
This is in contrast to \cite{Hirano2000}, who specify a strongly parametric probability model for all four of the unidentified parameters.

We consider three priors on $\theta$. All three are of the form
\begin{equation}
(\mbox{logit}\{\gamma^{00}_n(\X)\}, \mbox{logit}\{\gamma^{11}_a(\X)\})' = \mu\{\phi(\X)\} + h(\X)\beta + \epsilon(x)\label{eq:theta-3}.
\end{equation}
with $\mu(\phi(\X)) = (\mbox{logit}\{\gamma^{01}_n(\X)\}, \mbox{logit}\{\gamma^{10}_a(\X)\})'$, $h(\X)=1$ and $\beta=(b_a, -b_n)$. This centers our prior at a proportional odds model, where the reminder increases the odds of a positive outcome by a factor of $\exp(b_a)$ or $\exp(b_n)$ in always or never takers (respectively) at any covariate value. A natural way to complete this prior is to take $b_s\sim N(0, \sigma_s^2)$ for $s=a,n$ which centers the prior at the exclusion restrictions and expresses prior beliefs about plausible effect sizes through $\sigma_s^2$. For illustration we take $\sigma_s=0.1$ or $0.25$ for the current application. \cite{Hirano2000} estimate principal strata-independent total effects of COPD around -0.35 on the log odds scale, so the more diffuse prior covers a plausible range of possible effect sizes for the direct effect of a reminder within non-compliers.  We would expect any direct beneficial effect of the reminder on the subsequent hospitalization to be smaller than the detrimental effect of a serious chronic disease like COPD purely on substantive grounds.

In the flu application we also have some prior information about likely effect directions, which we can also incorporate in the prior. 
Any direct effect from the reminder is most likely to be positive, since the reminder probably encourages doctors and patients to take other preventative measures beyond the vaccine. Additionally, \cite{Hirano2000} speculate that the exclusion restriction for never takers is more plausible than for always takers since perceived risk is a key driver of compliance behavior, with low-risk patients more likely to be never takers. A more nuanced view is that the direct effect should be \emph{larger} for always takers than never takers --- even though never takers probably have lower risk on average, we should not rule out the possibility of high-risk never takers who refuse the vaccine due to cost, safety or other concerns. 

A prior for $\theta$ that reflects these beliefs is $(b_a, b_n)\sim \N(b_0, \Sigma_b)\mathbb{1}(b_a>0, b_n>0)$. For the analysis in the sequel we will take $b_0 = (0.05, 0.0)'$, $\Sigma_b = \mbox{diag}(0.13^2, 0.05^2)$. In additional simulations we find that results are insensitive to plausible values of $\mbox{Cov}(b_a, b_n)$. These particular numerical values were chosen  The prior mode is $b_0$, the prior mean is about $(0.125, 0.04)$, and the marginal $90^{th}$ percentile is about 0.25 for $b_a$ and 0.08 for $b_n$. 
The prior scale for $b_a$ is calibrated so that (with high probability) $b_a$ is smaller in magnitude than the effect of COPD on the hospitalization as reported by \cite{Hirano2000} (the posterior mean was generally around -0.35, SD 0.15). We will call this the ``informed" prior.
For all three priors we set $v=0.025$. 
\subsection{Results}

\begin{figure}
\begin{center}
\includegraphics[width=5in]{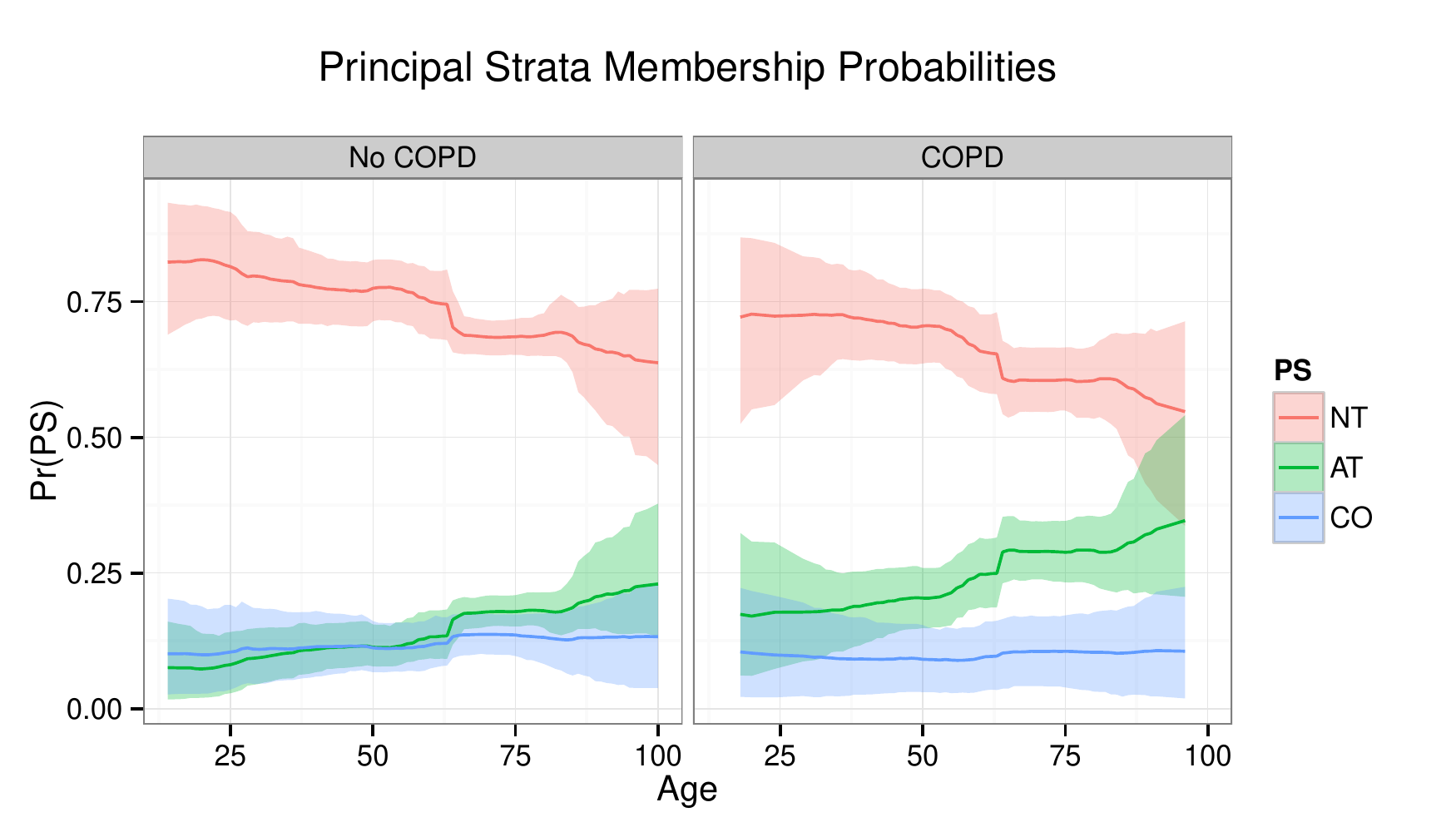}
\caption{Principal strata memberhsip probabilities by age and COPD, including marginal 90\% credible intervals}\label{fig:psmembership}
\end{center}
\end{figure}

We fit the model described in the previous section to the flu vaccine data, using age and COPD as covariates. Figure \ref{fig:psmembership} shows the probability of belonging to each principal stratum as a function of age and COPD along with marginal 90\% credible intervals. We find a fairly sharp break in compliance behavior at age 65 (the probability of being a never-taker drops, and seems to be shifted largely to always takers), which appears to hold regardless of COPD status. Possible explanations include patients or providers giving more weight to age-based vaccination guidelines, or better insurance coverage and increased access to services in the older cohort. 

\begin{figure}
\begin{center}
\includegraphics[width=3.5in]{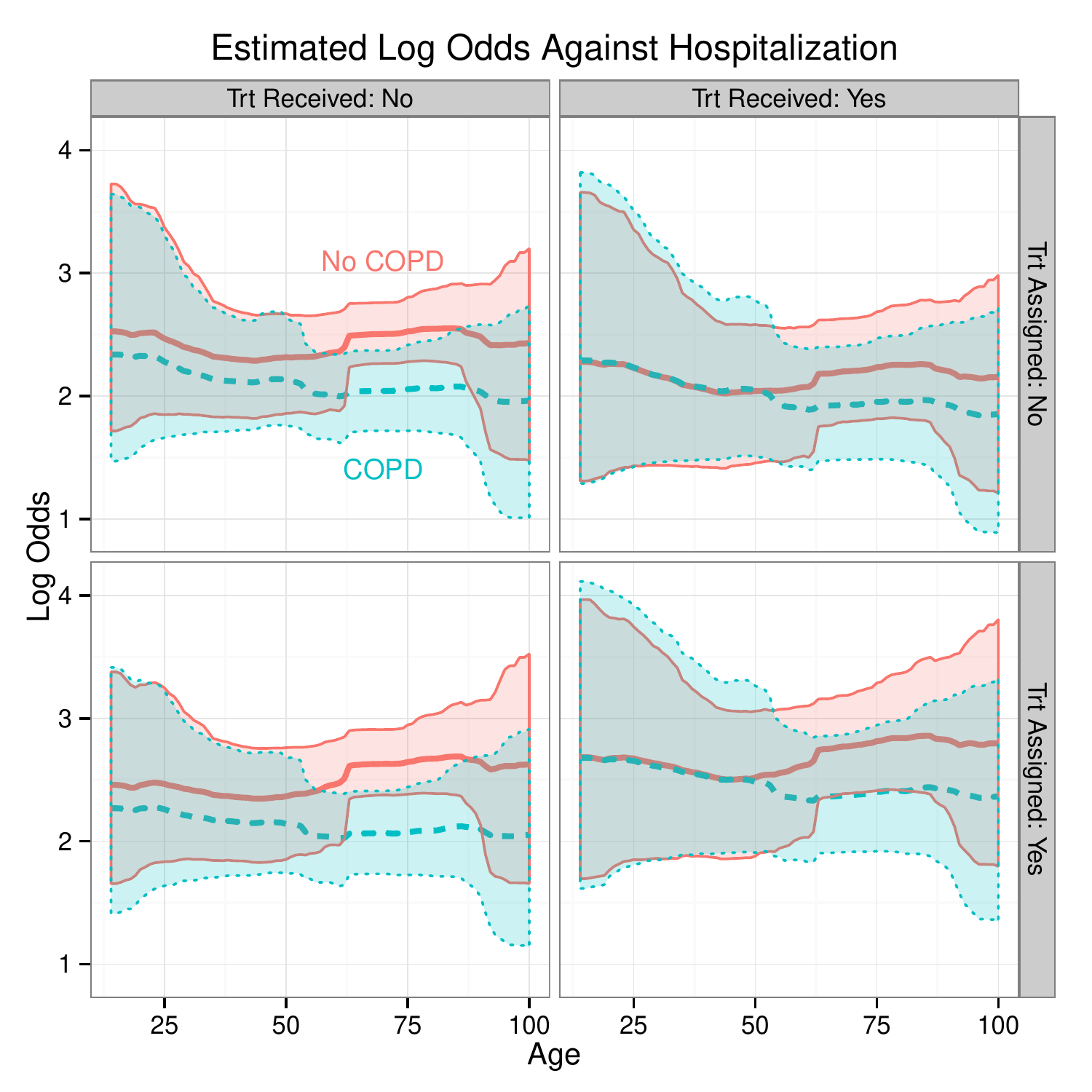}
\caption{Log odds of a positive outcome (no hospitalization) by age, COPD, and treatment assignment/receipt}\label{fig:outcometz}
\end{center}
\end{figure}

Figure \ref{fig:outcometz} shows the outcome distributions by age, COPD, and the treatment assigned and received. There is clear evidence of an age/COPD interaction; this is due at least in part to the study design, where all participants under 65 have at least one chronic disease (possibly COPD). This also means that the 65+ population is on average less likely to he hospitalized than the younger cohort, explaining the seemingly paradoxical decrease in the estimated probability of hospitalization at age 65 in the ``No COPD" group.

\subsubsection{Causal effects}
We are primarily interested in (partially identified) sample causal effects. 
As in \cite{Hirano2000} we will consider estimating the intention-to-treat (ITT) effect in each principal stratum:
\begin{equation}
ITT_s = \frac{1}{n_s}\sum_{i: S_i=s} (Y_i(T_i(1), 1) - Y_i(T_i(0), 0)). 
\end{equation}
where $n_s$ is the number of observations with compliance type $S$. Given a sample from the marginal distribution for $\phi$, and a sample from $\theta$, we can impute each observation's missing principal strata membership and the unobserved potential outcome. With the completed data it is straightforward to compute the sample ITT effects. Computational details are provided in Appendix \ref{app:postmis}


%

We compare results under our three priors to three models from \cite{Hirano2000}: the model with no exclusion restrictions (``none"), both exclusion restrictions (``both"), or the never-taker exclusion restriction (``n"). The ``Hirano (none)" model has very wide standard deviations, which are driven by their use of diffuse, ``off-the-shelf" priors.  Their marginal prior for the direct effect of the reminder is approximately normal with mean zero and standard deviation 1.5, which puts significant prior probability on implausible values such as large and/or negative direct effects. Such vague priors are inappropriate for partially identified parameters.

In contrast, the BART models all give similar results. The ``BART (0.1)" and ``BART (0.25)" models differ mainly in the width of the posterior standard errors, and then only slightly. The ``BART (0.25)" model assumes that with high probability the magnitude of any direct effect on the log odds is within $\pm 0.5$ -- compare this to the COPD effect, which is estimated to increase the log odds of hospitalization by about 0.35. The ``informed" prior restricts the population effects to be positive but small, which tends to shrink the $ITT_c$ effect toward zero -- the overall $ITT$ effect is identified, so positive effects in the other principal strata lead necessarily to smaller $ITT_c$ estimates than unrestricted models.

The substantive conclusion is largely unchanged: there is only modest evidence for the effectiveness of the vaccine, particularly if we assume the direct effect of the reminder is similar in always-takers and compliers. This is unsurprising given the sample size of $n=2,861$, weakness of the encouragement (Fig. \ref{fig:psmembership}), and relative rarity of the outcome (about 8\% of the study participants were hospitalized). However, unlike \cite{Hirano2000} we conclude that the $ITT_c$ effect is actually fairly robust to plausible deviations from the exclusion restrictions.


\begin{figure}
\begin{center} 
\includegraphics[width=5.25in]{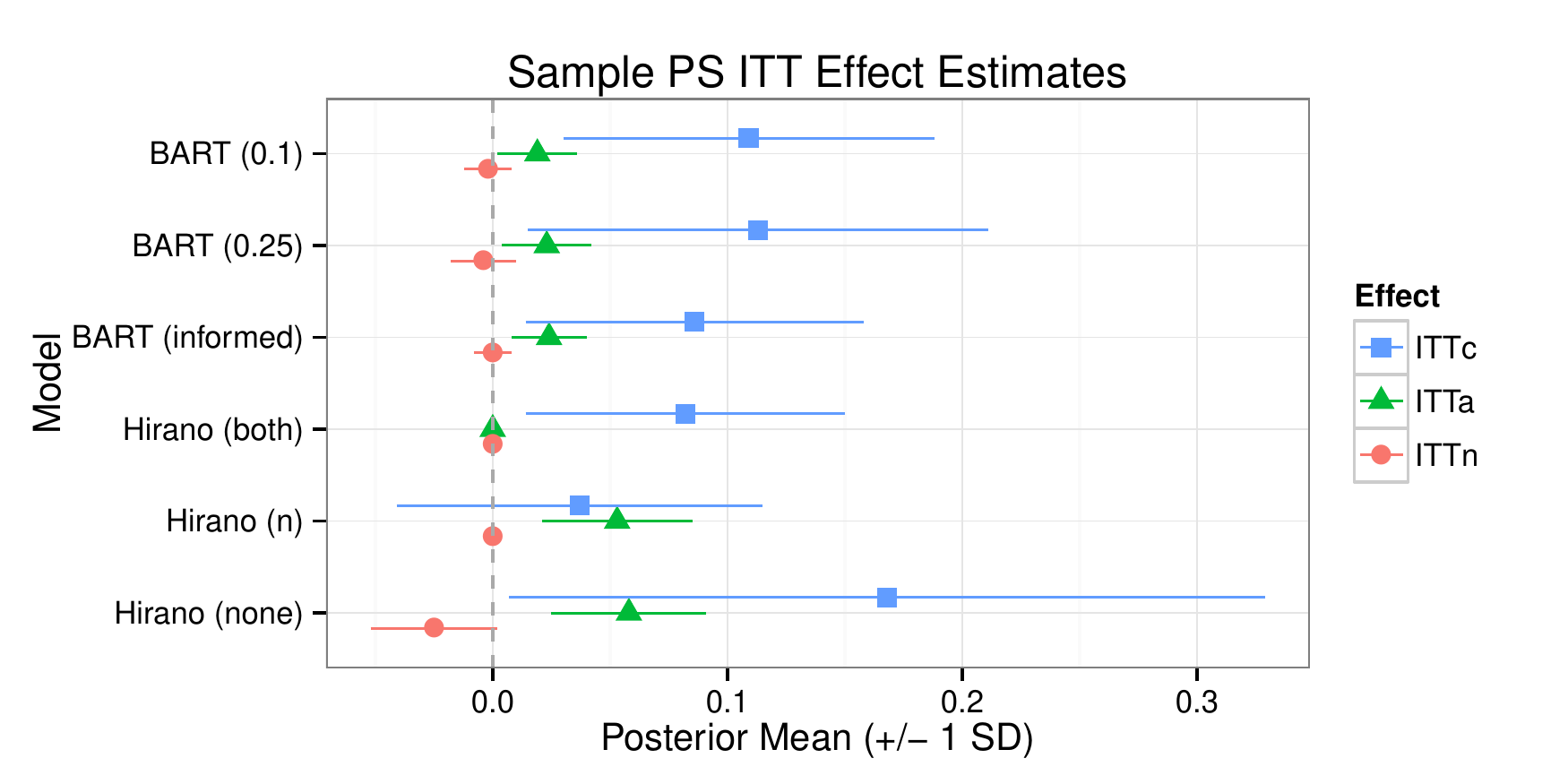}
\caption{Posterior mean and standard deviations for the sample ITT effects in each principal stratum. ``BART (0.1)" and ``BART (0.25)" use the conditional prior centered at the exclusion restrictions with $\sigma_a=\sigma_n=0.1$ or $0.25$, respectively. ``BART (informed)" includes all prior information. The ``Hirano" models are from \cite{Hirano2000}; the ``n" model assumes the exclusion restriction in never-takers, the ``both" model imposes both exclusion restrictions and the ``none" model has no exclusion restrictions.}\label{fig:flures}  
\end{center}
\end{figure}


 

\section{Imputing principal strata membership, potential outcomes, and $ITT$ effects}\label{app:postmis}

The full conditional distribution of principal strata membership is given by
\begin{align}
\Pr(S_i=a\mid Y_i=y, T_i=t, Z_i=z, X_i=x, \phi, \theta) &\propto \pi_A(x)\gamma_{a}^{11}(x)^{y}(1-\gamma_{a}^{11}(x))^{1-y}\mathbb{1}(Z_i=T_i=1)\\
\Pr(S_i=n\mid Y_i=y, T_i=t, Z_i=z, X_i=x, \phi, \theta) &\propto \pi_N(x)\gamma_{n}^{00}(x)^{y}(1-\gamma_{n}^{00}(x))^{1-y}\mathbb{1}(Z_i=T_i=0)\\
\Pr(S_i=c\mid Y_i=y, T_i=t, Z_i=z, X_i=x, \phi, \theta) &\propto \pi_C(x)\gamma_{c}^{tz}(x)^{y}(1-\gamma_{c}^{tz}(x))^{1-y}\mathbb{1}(Z_i=T_i)
\end{align}
Note that observations with $Z_i=1, T_i=0$ ($Z_i=0,\ T_i=1)$ are never takers (always takers) with probability one, and observations with $T_i=Z_i=1$ ($T_i=Z_i=0$) are either compliers or always takers (never takers). 

Finally, given an observation's PS membership, the missing potential outcome is sampled from its full conditional, which is just
\[
\Pr(Y_i(t, 1-z)=1\mid T_i=t, Z_i=z, X_i=x, S_i=s, \phi, \theta) = \gamma_s^{t(1-z)}(x) .
\]
The $ITT$ effects can be computed from the completed data using the following alternative expressions:
\begin{align}
ITT_c &= \frac{
\sum_{i=1}^n \mathbb{1}(S_i=c)[Z_i\{Y_i-Y_i(0,0)\}+ (1-Z_i)\{Y_i(1,1) - Y_i\}]
}{
\sum_{i=1}^n \mathbb{1}(S_i=c)
}\\
ITT_n &= \frac{
\sum_{i=1}^n \mathbb{1}(S_i=n)[Z_i\{Y_i-Y_i(0,0)\}+ (1-Z_i)\{Y_i(0,1) - Y_i\}]
}{
\sum_{i=1}^n \mathbb{1}(S_i=n)
}\\
ITT_c &= \frac{
\sum_{i=1}^n \mathbb{1}(S_i=a)[Z_i\{Y_i-Y_i(1,0)\}+ (1-Z_i)\{Y_i(1,1) - Y_i\}]
}{
\sum_{i=1}^n \mathbb{1}(S_i=a)
}
\end{align}

\singlespace 
\bibliographystyle{plainnat} 
\bibliography{bib/isolating_causalB}

\begin{thebibliography}{59}
\providecommand{\natexlab}[1]{#1}
\providecommand{\url}[1]{\texttt{#1}}
\expandafter\ifx\csname urlstyle\endcsname\relax
  \providecommand{\doi}[1]{doi: #1}\else
  \providecommand{\doi}{doi: \begingroup \urlstyle{rm}\Url}\fi

\bibitem[SEC(2014)]{SECwebsite}
{Securities and Exchange Commission, Accounting and Auditing Enforcement
  Releases}.
\newblock \url{http://www.sec.gov/divisions/enforce/friactions.shtml}, 2014.

\bibitem[Achen and Shively(1995)]{AchenShively199505}
C.H Achen and W.~Phillips Shively.
\newblock \emph{Cross-Level Inference}.
\newblock University Of Chicago Press, first edition, 5 1995.

\bibitem[Aldrich(2002)]{Aldrich2002}
J.~Aldrich.
\newblock How likelihood and identification went {B}ayesian.
\newblock \emph{International Statistical Review}, 70\penalty0 (1):\penalty0
  79--98, 2002.

\bibitem[Angrist et~al.(1996)Angrist, Imbens, and Rubin]{Angrist1996}
J.D. Angrist, G.W. Imbens, and D.B. Rubin.
\newblock Identification of causal effects using instrumental variables.
\newblock \emph{Journal of the American statistical Association}, 91\penalty0
  (434):\penalty0 444--455, 1996.

\bibitem[Barnard et~al.(2003)Barnard, Frangakis, Hill, and
  Rubin]{barnard2003principal}
J.~Barnard, C.E. Frangakis, J.L. Hill, and D.B. Rubin.
\newblock {Principal stratification approach to broken randomized experiments:
  A case study of school choice vouchers in New York City}.
\newblock \emph{Journal of the American Statistical Association}, 98\penalty0
  (462):\penalty0 299--323, 2003.

\bibitem[Chan and Tobias(2014)]{Chan2014}
J.C.C. Chan and J.L. Tobias.
\newblock Priors and posterior computation in linear endogenous variable models
  with imperfect instruments.
\newblock \emph{Journal of Applied Econometrics}, 2014.
\newblock ISSN 1099-1255.

\bibitem[Chipman et~al.(1998)Chipman, George, and
  McCulloch]{chipman1998bayesian}
H.A. Chipman, E.I. George, and R.E. McCulloch.
\newblock Bayesian {CART} model search.
\newblock \emph{Journal of the American Statistical Association}, 93\penalty0
  (443):\penalty0 935--948, 1998.

\bibitem[Chipman et~al.(2010)Chipman, George, and McCulloch]{chipman2010bart}
H.A. Chipman, E.I. George, and R.E. McCulloch.
\newblock {BART}: {B}ayesian additive regression trees.
\newblock \emph{The Annals of Applied Statistics}, 4\penalty0 (1):\penalty0
  266--298, 2010.

\bibitem[Cox(1958)]{cox1958regression}
D.R. Cox.
\newblock The regression analysis of binary sequences.
\newblock \emph{Journal of the Royal Statistical Society. Series {B}
  (Methodological)}, pages 215--242, 1958.

\bibitem[Daniels and Hogan(2008)]{Daniels2008}
M.J. Daniels and J.W. Hogan.
\newblock \emph{Missing data in longitudinal studies: Strategies for Bayesian
  modeling and sensitivity analysis}.
\newblock CRC Press, 2008.

\bibitem[Dawid(1979)]{Dawid1979}
A.P. Dawid.
\newblock Conditional independence in statistical theory.
\newblock \emph{Journal of the Royal Statistical Society. Series B},
  41\penalty0 (1):\penalty0 1--31, 1979.

\bibitem[Dechow et~al.(2011)Dechow, Ge, Larson, and
  Sloan]{dechow2011predicting}
P.M. Dechow, W.~Ge, C.R. Larson, and R.G. Sloan.
\newblock Predicting material accounting misstatements.
\newblock \emph{Contemporary accounting research}, 28\penalty0 (1):\penalty0
  17--82, 2011.

\bibitem[Duncan and Davis(1953)]{duncan1953alternative}
O.D. Duncan and B.~Davis.
\newblock An alternative to ecological correlation.
\newblock \emph{American Sociological Review}, 18\penalty0 (6):\penalty0
  665--666, 1953.

\bibitem[Dyck et~al.(2008)Dyck, Volchkova, and Zingales]{dyck2008corporate}
A.~Dyck, N.~Volchkova, and L.~Zingales.
\newblock The corporate governance role of the media: {E}vidence from {R}ussia.
\newblock \emph{The Journal of Finance}, 63\penalty0 (3):\penalty0 1093--1135,
  2008.

\bibitem[Dyck et~al.(2013)Dyck, Morse, and Zingales]{Dyck2013}
A.~Dyck, A.~Morse, and L.~Zingales.
\newblock How pervasive is corporate fraud?
\newblock Technical Report 2222608, Rotman School of Management, 2013.

\bibitem[Florens and Simoni(2011)]{florens2011bayesian}
J.-P. Florens and A.~Simoni.
\newblock Bayesian identification and partial identification.
\newblock \emph{Unpublished manuscript. Toulouse School of Economics, Toulouse,
  France}, 2011.

\bibitem[Fr{\'e}chet(1951)]{frechet1951tableaux}
M.~Fr{\'e}chet.
\newblock Sur les tableaux de corr{\'e}lation dont les marges sont donn{\'e}es.
\newblock \emph{Ann. Univ. Lyon Sect. A}, 9:\penalty0 53--77, 1951.

\bibitem[Frisch(1934)]{frisch1934statistical}
R.~Frisch.
\newblock \emph{Statistical Confluence Analysis by Means of Complete Regression
  Systems}.
\newblock Universitetets {\O}konomiske Instituut, 1934.

\bibitem[Gelfand and Sahu(1999)]{GelfandSahu1999}
A.E. Gelfand and S.K. Sahu.
\newblock Identifiability, improper priors and {G}ibbs sampling for generalized
  linear models.
\newblock \emph{Journal of the American Statistical Association}, 94\penalty0
  (445):\penalty0 247--253, 1999.

\bibitem[Greenland(2005)]{greenland2005multiple}
S.~Greenland.
\newblock Multiple-bias modelling for analysis of observational data.
\newblock \emph{Journal of the Royal Statistical Society: Series A (Statistics
  in Society)}, 168\penalty0 (2):\penalty0 267--306, 2005.

\bibitem[Gustafson(2010)]{gustafson2010bayesian}
P.~Gustafson.
\newblock Bayesian inference for partially identified models.
\newblock \emph{The International Journal of Biostatistics}, 6\penalty0 (2),
  2010.

\bibitem[Gustafson et~al.(2005)Gustafson, Gelfand, Sahu, Johnson, Hanson,
  Joseph, and Lee]{gustafson2005model}
P.~Gustafson, A.E. Gelfand, S.K. Sahu, W.O. Johnson, T.E. Hanson, L.~Joseph,
  and J.~Lee.
\newblock On model expansion, model contraction, identifiability and prior
  information: Two illustrative scenarios involving mismeasured variables [with
  comments and rejoinder].
\newblock \emph{Statistical Science}, pages 111--140, 2005.

\bibitem[Haavelmo(1943)]{haavelmo1943}
T.~Haavelmo.
\newblock The statistical implications of a system of simultaneous equations.
\newblock \emph{Econometrica}, 11\penalty0 (1):\penalty0 pp. 1--12, 1943.

\bibitem[Haavelmo(1944)]{haavelmo1944}
T.~Haavelmo.
\newblock The probability approach in econometrics.
\newblock \emph{Supplement to Econometrica}, 12:\penalty0 iii--115, July 1944.

\bibitem[Heckman(1976)]{Heckman1976}
J.J. Heckman.
\newblock Simultaneous equation models with continuous and discrete endogenous
  variables and structural shifts.
\newblock In S.M. Goldfeld and R.E. Quandt, editors, \emph{Studies in nonlinear
  estimation}, pages 235--272. Balinger, 1976.

\bibitem[Heckman(1978)]{Heckman1978}
J.J. Heckman.
\newblock Dummy endogenous variables in a simultaneous equation system.
\newblock \emph{Econometrica}, 46:\penalty0 931--959, 1978.

\bibitem[Heckman(1979)]{Heckman1979}
J.J. Heckman.
\newblock Sample bias as a specification error.
\newblock \emph{Econometrica}, 47:\penalty0 153--161, 1979.

\bibitem[Hill(2012)]{Hill2012}
J.L. Hill.
\newblock Bayesian nonparametric modeling for causal inference.
\newblock \emph{Journal of Computational and Graphical Statistics}, 20\penalty0
  (1):\penalty0 212--240, 2012.

\bibitem[Hirano et~al.(2000)Hirano, Imbens, Rubin, and Zhou]{Hirano2000}
K.~Hirano, G.W. Imbens, D.B. Rubin, and X.H. Zhou.
\newblock {Assessing the effect of an influenza vaccine in an encouragement
  design.}
\newblock \emph{Biostatistics}, 1\penalty0 (1):\penalty0 69--88, March 2000.

\bibitem[Imbens and Rubin(1997{\natexlab{a}})]{Imbens1997a}
G.~W. Imbens and D.~B. Rubin.
\newblock {Estimating Outcome Distributions for Compliers in Instrumental
  Variables Models}.
\newblock \emph{The Review of Economic Studies}, 64\penalty0 (4):\penalty0
  555--574, October 1997{\natexlab{a}}.
\newblock ISSN 0034-6527.
\newblock \doi{10.2307/2971731}.
\newblock URL
  \url{http://restud.oxfordjournals.org/lookup/doi/10.2307/2971731}.

\bibitem[Imbens and Rubin(1997{\natexlab{b}})]{imbens1997bayesian}
G.W. Imbens and D.B. Rubin.
\newblock Bayesian inference for causal effects in randomized experiments with
  noncompliance.
\newblock \emph{The Annals of Statistics}, pages 305--327, 1997{\natexlab{b}}.

\bibitem[Jayaraman and Milbourn(2010)]{jayaraman2010whistle}
S.~Jayaraman and T.~Milbourn.
\newblock {Whistle Blowing and CEO Compensation: The Qui Tam Statute}.
\newblock In \emph{AFA 2011 Denver Meetings Paper}, 2010.

\bibitem[Kadane(1975)]{Kadane1975}
J.~Kadane.
\newblock The role of identification in {B}ayesian theory.
\newblock In S.E. Fienberg and A.~Zellner, editors, \emph{Studies in {B}ayesian
  econometrics and statistics}, chapter 5.2, pages 175--191. North-Holland,
  1975.

\bibitem[King(2013)]{king2013solution}
G.~King.
\newblock \emph{A solution to the ecological inference problem: Reconstructing
  individual behavior from aggregate data}.
\newblock Princeton University Press, 2013.

\bibitem[Kline and Tamer(2013)]{kline2013default}
B.~Kline and E.~Tamer.
\newblock Default {B}ayesian inference in a class of partially identified
  models.
\newblock \emph{manuscript, Northwestern University}, 2013.

\bibitem[Koopmans(1949)]{Koopmans1949}
T.C. Koopmans.
\newblock Identification problems in economic model construction.
\newblock \emph{Econometrica}, pages 125--144, 1949.

\bibitem[Koopmans and Reiersol(1950)]{KoopmansReiersol1950}
T.C. Koopmans and O.~Reiersol.
\newblock The identification of structural characteristics.
\newblock \emph{Annals of Mathematical Statistics}, 21\penalty0 (2):\penalty0
  165--181, 1950.

\bibitem[Lancaster and Imbens(1996)]{Lancaster1996}
T.~Lancaster and G.~Imbens.
\newblock Case-control studies with contaminated controls.
\newblock \emph{Journal of Econometrics}, 71:\penalty0 145--160, 1996.

\bibitem[Little and Rubin(2002)]{little1987statistical}
R.J.A. Little and D.B. Rubin.
\newblock \emph{Statistical Analysis with Missing Data}.
\newblock Wiley-Interscience, 2 edition, 9 2002.
\newblock ISBN 9780471183860.

\bibitem[Manski(1995)]{manski1995identification}
C.F. Manski.
\newblock \emph{Identification problems in the social sciences}.
\newblock Harvard University Press, 1995.

\bibitem[Manski(2003)]{manski2003partial}
C.F. Manski.
\newblock \emph{Partial identification of probability distributions}.
\newblock Springer, 2003.

\bibitem[Manski(2007)]{manski2007identification}
C.F. Manski.
\newblock \emph{Identification for prediction and decision}.
\newblock Harvard University Press, 2007.

\bibitem[McCandless et~al.(2007)McCandless, Gustafson, and
  Levy]{McCandless2007}
L.C. McCandless, P.~Gustafson, and A.~Levy.
\newblock {B}ayesian sensitivity analysis for unmeasured confounding in
  observational studies.
\newblock \emph{Statistics in medicine}, 26\penalty0 (11):\penalty0 2331--2347,
  2007.

\bibitem[McCandless et~al.(2012)McCandless, Gustafson, Levy, and
  Richardson]{McCandless2012}
L.C. McCandless, P.~Gustafson, A.R. Levy, and S.~Richardson.
\newblock Hierarchical priors for bias parameters in {B}ayesian sensitivity
  analysis for unmeasured confounding.
\newblock \emph{Statistics in medicine}, 31\penalty0 (4):\penalty0 383--396,
  2012.

\bibitem[McDonald et~al.(1991)McDonald, Hui, and Tierney]{McDonald1992}
C.J. McDonald, S.L. Hui, and W.M. Tierney.
\newblock Effects of computer reminders for influenza vaccination on morbidity
  during influenza epidemics.
\newblock \emph{{MD computing: computers in medical practice}}, 9\penalty0
  (5):\penalty0 304--312, 1991.

\bibitem[Miller(2006)]{miller2006press}
Gregory~S Miller.
\newblock The press as a watchdog for accounting fraud.
\newblock \emph{Journal of Accounting Research}, 44\penalty0 (5):\penalty0
  1001--1033, 2006.

\bibitem[Molitor et~al.(2009)Molitor, Best, Jackson, and
  Richardson]{Molitor2009}
N.-T. Molitor, N.~Best, C.~Jackson, and S.~Richardson.
\newblock Using {B}ayesian graphical models to model biases in observational
  studies and to combine multiple sources of data: application to low birth
  weight and water disinfection by-products.
\newblock \emph{Journal of the Royal Statistical Society: Series A (Statistics
  in Society)}, 172\penalty0 (3):\penalty0 615--637, 2009.

\bibitem[Moon and Schorfheide(2012)]{moon2012bayesian}
H.R. Moon and F.~Schorfheide.
\newblock Bayesian and frequentist inference in partially identified models.
\newblock \emph{Econometrica}, 80\penalty0 (2):\penalty0 755--782, 2012.

\bibitem[Pakman and Paninski(2014)]{HMC}
Ari Pakman and Liam Paninski.
\newblock Exact hamiltonian monte carlo for truncated multivariate gaussians.
\newblock \emph{Journal of Computational and Graphical Statistics}, 23\penalty0
  (2):\penalty0 518--542, 2014.

\bibitem[Phillips~Shively(1969)]{shively1969ecological}
W.~Phillips~Shively.
\newblock {``Ecological" inference: The use of aggregate data to study
  individuals}.
\newblock \emph{The American Political Science Review}, pages 1183--1196, 1969.

\bibitem[Poirier(1980)]{poirier1980partial}
D.J. Poirier.
\newblock Partial observability in bivariate probit models.
\newblock \emph{Journal of Econometrics}, 12\penalty0 (2):\penalty0 209--217,
  1980.

\bibitem[Prentice and Pyke(1979)]{prentice1979logistic}
R.L. Prentice and R.~Pyke.
\newblock Logistic disease incidence models and case-control studies.
\newblock \emph{Biometrika}, 66\penalty0 (3):\penalty0 403--411, 1979.

\bibitem[Richardson and Robins(2010)]{Richardson2010}
T.S. Richardson and J.M. Robins.
\newblock {Analysis of the binary instrumental variable model}.
\newblock In Rina Dechter, Hector Geffner, and Joseph~Y. Halpern, editors,
  \emph{Heuristics, Probability and Causality}, chapter~25. College
  Publications, 2010.

\bibitem[Richardson et~al.(2011)Richardson, Evans, and Robins]{Richardson2011}
T.S. Richardson, R.J. Evans, and J.M. Robins.
\newblock {Transparent parameterizations of models for potential outcomes}.
\newblock In J.M. Bernardo, M.J. Bayarri, James~O. Berger, A.~P. Dawid, David
  Heckerman, A.F.M Smith, and Mike West, editors, \emph{Bayesian Statistics},
  volume~9, pages 569--610. Oxford University Press, 2011.

\bibitem[Rubin(1976)]{rubin1976inference}
D.B. Rubin.
\newblock Inference and missing data.
\newblock \emph{Biometrika}, 63\penalty0 (3):\penalty0 581--592, 1976.

\bibitem[San~Mart\'{i}n and Gonz\'{a}lez(2010)]{MartinGonzalez2010}
E.~San~Mart\'{i}n and J.~Gonz\'{a}lez.
\newblock {B}ayesian identifiability: contributions to an inconclusive debate.
\newblock \emph{Chilean Journal of Statistics}, 1\penalty0 (2):\penalty0
  69--91, 2010.

\bibitem[Tamer(2010)]{Tamer2010}
E.~Tamer.
\newblock Partial identification in econometrics.
\newblock \emph{Annual Review of Economics}, 2:\penalty0 167--95, 2010.

\bibitem[Wang(2013)]{Wang2013}
T.Y. Wang.
\newblock Corporate securities fraud: insights from a new empirical framework.
\newblock \emph{Journal of Law, Economics and Organization}, 29\penalty0
  (3):\penalty0 535--568, 2013.

\bibitem[Zhang and Rubin(2003)]{zhang2003estimation}
J.L. Zhang and D.B. Rubin.
\newblock Estimation of causal effects via principal stratification when some
  outcomes are truncated by ``death''.
\newblock \emph{Journal of Educational and Behavioral Statistics}, 28\penalty0
  (4):\penalty0 353--368, 2003.

\end{thebibliography}

%

\end{document}